\documentclass[12pt]{article}
\usepackage{amsmath}
\usepackage{graphicx}
\usepackage{enumerate}
\usepackage{natbib}
\usepackage{array}
\usepackage{algorithmic}
\usepackage[plain]{algorithm}
\usepackage{booktabs,tabulary,array,tabularx}
\usepackage{siunitx}
\usepackage{threeparttable}
\DeclareMathOperator*{\plim}{plim}

\newcommand*{\doublerule}{\hrule width \hsize height 0.5pt \kern 0.5mm \hrule width \hsize height 0.5pt}
\newcommand*{\singlerule}{\hrule width \hsize height 0.5pt}
\usepackage[colorlinks = true,
linkcolor = red,
urlcolor  = blue,
filecolor={red},
citecolor = blue,
anchorcolor = blue]{hyperref}
\usepackage{xurl}
\newcommand{\blind}{1}
\usepackage{mathtools}
\usepackage{etoolbox}
\usepackage[left=2.5cm, right=2.5cm, top=2.5cm]{geometry}
\newcommand{\proofend}{$\hfill\Box{~}$}
\newenvironment{Proof}{\noindent {\em{\bf Proof.}}}{\proofend\\}
\usepackage{xcolor}

\usepackage{caption}
\usepackage{graphicx,latexsym,amssymb,amsmath}
\usepackage{subcaption}
\usepackage{bbm, bm}

\def\R{{\mathbb{R}}}

\def\P{{\mathbb{P}}}
\def\E{{\mathbb{E}}}
\usepackage{newfloat}
\usepackage{pdflscape}
\usepackage{afterpage}
\usepackage{capt-of}
\newtheorem{Lemma}{Lemma}[section] 
\newtheorem{Assumption}{Assumption}
\newtheorem{Theorem}{Theorem}

\begin{document}

\def\spacingset#1{\renewcommand{\baselinestretch}%
{#1}\small\normalsize} \spacingset{1}


\if1\blind
{
  \title{\bf Counterfactuals in factor models}
  \author{Jad Beyhum\hspace{.2cm}\\
     Department of Economics, KU Leuven, Belgium}
  \maketitle
} \fi

\if0\blind
{
  \bigskip
  \bigskip
  \bigskip
  \begin{center}
    {\LARGE\bf Counterfactuals in factor models}
\end{center}
  \medskip
} \fi

\bigskip
\begin{abstract}
We study a new model where the potential outcomes, corresponding to the values of a (possibly continuous) treatment, are linked through common factors. The factors can be estimated using a panel of regressors. We propose a procedure to estimate time-specific and unit-specific average marginal effects in this context. Our approach can be used either with high-dimensional time series or with large panels. It allows for treatment effects heterogenous across time and units and is straightforward to implement since it only relies on principal components analysis and elementary computations. We derive the asymptotic distribution of our estimator of the average marginal effect and highlight its solid finite sample performance through a simulation exercise. The approach can also be used to estimate average counterfactuals or adapted to an instrumental variables setting and we discuss these extensions. Finally, we illustrate our novel methodology through an empirical application on income inequality.
\end{abstract}

\noindent%

\medskip

\noindent%
\vfill
\newpage
\spacingset{1.5} 

\section{Introduction} The goal is to infer the causal effect of a possibly continuous treatment variable $d_{it}\in\R$ on an outcome variable in a panel data setting. Let $y_{it}(d)$ be the potential outcome  under treatment status $d$ of unit $i\in\{1,\dots,N\}$ at date $t\in\{1,\dots,T\}$. The realized outcome of unit $i$ at time $t$ is given by $y_{it}=y_{it}(d_{it})$. Our main parameters of interest are the unit-specific and time-specific average marginal effects (AME) of the treatment. The unit-specific AME of unit $i$ is
$$\Delta_i =\E_t[y_{it}'(d_{it})],$$
where $y_{it}'(\cdot)$ is the derivative of $y_{it}(\cdot)$ and $\E_t$  denotes the expectation over $t$, for fixed $i$ formally, $\E_t[y_{it}'(d_{it})]=\plim\limits_{T\to\infty}T^{-1}\sum_{t=1}^T y_{it}'(d_{it}) $. Note that, when we introduce derivatives or probability limits, we implicitly assume that they exist. Similarly, the time-specific AME at date $t$ is defined by 
$$\Delta_t=\E_i[y_{it}'(d_{it})],$$
where  $\E_i$  denotes the expectation over $i$, for fixed $t$, formally, $\E_i[y_{it}'(d_{it})]=\plim\limits_{N\to\infty}N^{-1}\sum_{i=1}^N y_{it}'(d_{it}).$ We are also interested in the AME over the whole population, that is $$\Delta=\E[\Delta_i]=\E[y_{it}'(d_{it})].$$ Other important parameters such as average counterfactuals and treatment effects are also discussed in the paper.

To estimate the AMEs, we assume that the potential outcomes are related through a set of random factors $f_t\in\R^R$ such that 
\begin{equation}\label{modelY} 
y_{it}(d)=\lambda_{*i}(d)^\top f_t+\tilde u_{it}(d),\ d\in\R,\ t=1,\dots,T.
\end{equation}
where $R$ is the number of factors, $\lambda_{i*}(d)\in\R^R$ is a vector of nonrandom loadings and $\tilde u_{it}(d)\in\R$ is a random error term. We have \begin{equation} \label{AME2} \Delta_i = \E_t[\lambda_{*i}'(d_{it})^\top f_t],\ \Delta_t = \E_i[\lambda_{*i}'(d_{it})^\top f_t],\ \Delta= \E[\lambda_{*i}'(d_{it})^\top f_t],\end{equation}
where $\lambda_{*i}'(\cdot)$ is the derivative of $\lambda_{*i}(\cdot)$, so that knowledge of $\lambda_{i*}(\cdot)$ and $f_t$ is sufficient to identify $\Delta$. 
When $d_{it}$ is continuous, we only observe $y_{it}(d)$ for each $d$ with probability $0$, so that $\lambda_{i*}(\cdot)$ and $f_t$ cannot be estimated from data on $y_{it}$. To overcome this issue, we leverage a panel $\{x_{\ell t}\in\R, \ell=1,\dots,L,\ t=1,\dots,T\}$ that loads on the factors. That is, there exists (deterministic) loadings $\lambda_{\ell}\in\R^R$ and random error terms $e_{\ell t}\in\R$ such that
\begin{equation}\label{modelX} 
x_{\ell t}=\lambda_{\ell}^\top f_t+e_{\ell t},\ \ \ell=1,\dots,L,\ t=1,\dots,T,
\end{equation}
the factors can then be estimated from this panel using principal components analysis (or any other method such as cross-section averages). The loadings $\lambda_{*i}(\cdot)$ can then be estimated by linear regression of $y_{it}$ on the estimated factors interacted with powers of $d_{it}$ and additional controls.

The proposed approach allows for treatment effects heterogenous across $i$ and $t$ and leverages the very common modelling assumption of factor models. This set-up can model high-dimensional time series when $N=1$ and both $L$ (number of variables in the time series) and $T$ are large. For instance, in a macroeconomic application, $d_{1t}$ could correspond to the US federal funds rate, while $y_{1t}$ is the unemployment rate and the panel $\{x_{\ell t}\}_{\ell,t}$ corresponds to the well-known FRED-MD dataset of \cite{mccracken2016fred}. Another interesting context corresponding to our set-up is that of a large $N,T$ panel. For instance, suppose that we observe a set of $N$ countries at $T$ dates. On top of the outcome and the treatment, for each country we observe $K$ variables at each of the $T$ dates. These variables then constitute the panel $\{x_{\ell t}\}_{\ell, t}$, with $L=KN$. \\

\noindent \textbf{Contributions.} 
As mentioned above, we formulate an estimation procedure for $\Delta_i, \Delta_t$, and $\Delta$. Initially, we estimate the factors from the panel $\{x_{\ell t}\}_{\ell, t}$ using principal components analysis (PCA). Subsequently, we perform a linear regression of $y_{it}$ on the estimated factors interacted with powers of $d_{it}$ and additional controls. This approach enables the estimation of $\lambda_{i*}(\cdot)$ and facilitates the recovery of the AMEs through plug-in methods. We establish the asymptotic properties of the estimator $\widehat{\Delta}_i$ for $\Delta_i$ in the high-dimensional time series case, where $T, L \to \infty$ while $N$ can remain fixed. Additionally, we demonstrate the asymptotic normality of the estimators $\widehat{\Delta}_t$ of $\Delta_t$ and $\widehat{\Delta}$ of $\Delta$ in the large $N, T$ panel case, where $T,N,L\to \infty$. The finite sample performance of the estimators is evaluated through simulations. An empirical application on income inequality using a large $N,T$ panel illustrates our novel methodology.

This paper introduces a comprehensive approach to model counterfactuals using a factor model. While our focus in this work centers on estimating AMEs due to their importance in nonlinear panel data models \citep{fernandez2018fixed}, our methodology is versatile. It readily extends to estimating various average counterfactuals such as $\E_i[y_{it}(d)], \E_t[y_{it}(d)]$, or $\E[y_{it}(d)]$ for fixed $d$, as well as implementing instrumental variables estimation. Section \ref{sec.ext} delves into these extensions and provides corresponding estimators. To maintain conciseness, we refrain from explicitly deriving the asymptotic distributions of these additional estimators. However, similar proof techniques to those employed in establishing the asymptotic normality of the estimators of the average marginal effects can be applied.
\\

\noindent \textbf{Related literature.} 
Modeling potential outcomes of a binary treatment using a factor structure is a well-established approach in the literature \citep{gobillon2016regional,athey2021matrix,bai2021matrix,fan2022we}. These papers use a factor model for the untreated potential outcome but do not model the ``treated'' counterfactual. Specifically, \cite{athey2021matrix,bai2021matrix} delve into the realm of causal matrix completion, employing a panel denoted as $\{y_{it}\}_{i,t}$. Given the binary nature of the treatment in their study, they are able to observe numerous entries of the panel $\{y_{it}(0)\}_{i,t}$. By assuming a factor structure on $y_{it}(0)$, they successfully recover $y_{it}(0)$ for all pairs of $i$ and $t$ by effectively ``completing" the matrix using the information from $\{y_{it}(0)\}_{i,t}$. However, it is crucial to note that this strategy is not applicable in our current study, which involves a continuous treatment. In this scenario, with probability $1$, none of the entries in the matrix $\{y_{it}(d)\}_{i,t}$ for a given treatment level $d$ will be observed, making it impossible to successfully complete the matrix (the same type of issues would arise with the approaches of \cite{gobillon2016regional}, \cite{fan2022we}, since we would also almost never observe the untreated potential outcome). To address this challenge, we propose (i) assuming that the factors are shared across all potential outcomes and (ii) leveraging their learnability from the panel $\{x_{\ell t}\}_{\ell, t}$. 

Another related strand of literature is that of panel data models with interactive fixed effects \citep{pesaran2006estimation,bai2009panel,greenaway2012asymptotic}. In this set-up it is assumed that $y_{it}=\beta_d d_{it}+\beta_x^\top \tilde x_{it}+\lambda_i^\top f_t+\epsilon_{it}$, where $\tilde x_{it}$ is a set of controls, $\lambda_i$ are nonrandom loadings and $\epsilon_{it}$ is an error term. Similarly to us, some approaches also assume that $\tilde x_{it}$ follows a factor model with $f_t$ as common factors \citep{pesaran2006estimation,greenaway2012asymptotic}. 
Translated to potential outcomes, panel data models with interactive fixed effects assume that $y_{it}(d)=\beta_d d+\beta_x^\top \tilde x_{it}+\lambda_i^\top f_t+\epsilon_{it}.$ Comparing with our model, we see that in our case the effect of $d$ directly interacts with the factors. Thanks to that, the treatment effects can be heterogenous across time and units in our framework. Instead, in panel data models with interactive fixed effects, although it is possible to let $\beta$ depend on $i$ \citep[this is called the ``heterogenous slopes case'', see][]{pesaran2006estimation}, one cannot have treatment effects heterogenous across both dimensions.  Moreover, since this approach assumes that the regressors follow a factor model, we argue that it is more natural to also assume that the potential outcomes also load on the same factors and that $d$ affects the loadings as we do rather than assuming that the effect of $d$ on the potential outcomes is just linear. Note also that if one factor is a constant, our model directly generalizes the panel data models with interactive fixed effects of \citep{pesaran2006estimation,greenaway2012asymptotic}. 

Next, as mentioned earlier, our approach allows us to estimate treatment effects with high-dimensional time series. A leading approach to estimate treatment effects in time series is local projections \citep{jorda2023local}. Local projection is a linear regression of the outcome on the treatment and some controls and this approach has been extended to high dimensional data \citep{babii2022high,adamek2022local}. The standard local projection methodology does not allow for time-varying treatment effects. We improve on local projections in this dimension, at the cost of having to assume that the data follows an approximate factor model.   

Finally, this paper is related to the literature on characteristics-based factor model \citep{connor2007semiparametric,connor2012efficient,fan2016projected}. This literature considers a large N,T panel $\{\tilde y_{it}\}_{i,t}$ and assumes that $\tilde y_{it}$ follows a factor model. The loadings are modelled as nonparametric functions of some time-invariant characteristics $\tilde{x}_i$. The loadings and factors can they be estimated by either a semiparametric least squares approach or a projection strategy. The present paper differs from this strand of research in at least two respects. First, we consider a causal inference problem with counterfactuals. Second, in our case the loadings are functions of a time-varying variable $d_{it}$. Since the approach of the aforementioned papers crucially relies on the fact that the loadings are functions of time-invariant characterustics, we cannot build up on their estimation procedure. For this reason, we developed a new methodology, where the factors are first estimated from the panel $\{x_{\ell t}\}_{\ell,t}$. We note that \cite{fan2016projected} provide a theory where the loadings are estimated nonparametrically using series. This is in the spirit of our approach, but to simplify the exposition of the current paper, we do not let the number of series terms go to $\infty$ in theory. \\

 \noindent \textbf{Outline.} The paper is organized as follows. Section \ref{sec.est} introduces the estimation procedure. Then, we expose the asymptotic theory in Section \ref{sec.th}. Extensions are discussed in Section \ref{sec.ext}. Section \ref{sec.sim} contains simulations. The methodology is illustrated through an empirical application in Section \ref{sec.emp}. Finally, we conclude in Section \ref{sec.ccl}.\\

\noindent \textbf{Notation.} For an integer $n$, we let $[n]=\{1,\dots,n\}$. For a $n_1\times n_2$ matrix $A$, its $k^{th}$ singular value is $\sigma_k(A)$ and $A=\sum_{k=1}^{\min(n_1,n_2)}\sigma_k(A)u_k(A)v_k(A)^\top$ is the singular value decomposition of $A$, where $\left\{u_k\left(A\right)\right\}_{k=1}^{\min(n_1,n_2)}$ is a family of orthonormal vectors of $\mathbb{R}^{n_1}$ and $\left\{v_k\left(A\right)\right\}_{k=1}^{\min(n_1,n_2)}$ is a family of orthonormal vectors of $\mathbb{R}^{n_2}$. The operator norm is $\|A\|_{\text{op}}=\sigma_1(A)$ and the Euclidian norm is $\|A\|_2=\sqrt{\sum_{i=1}^{n_1}\sum_{j=1}^{n_2}A_{ij}^2}$. For an integer $n$, $0_n$ and $1_n$ are, respectively, the $n\times 1$ vectors with all coefficients equal to 0 and 1. Moreover, $I_n$ is the identity matrix of size $n$. For integers $n_1$ and $n_2$, $0_{n_1,n_2}$ and $1_{n_1,n_2}$ are, respectively, the $n_1\times n_2$ matrices with all coefficients equal to 0 and 1.  For a random variable $s_{it}$, we let $\E_i[s_{it}]=\plim\limits_{N\to\infty} \frac1N \sum_{i=1}^N s_{it}$ and $\E_t[s_{it}]=\plim\limits_{T\to\infty} \frac1T \sum_{t=1}^T s_{it}$, when the latter quantities exist.

\section{Estimation}\label{sec.est}
 In this paper, we estimate the factor by principal components analysis (PCA) applied to the $T\times L$ matrix X such that $X_{t\ell } = x_{\ell t}$. Note that \eqref{modelX} becomes
$$X=F\Lambda^\top+E$$ in matrix form, where $F=(f_1,\dots,f_T)^\top$ is $T\times R$, $\Lambda=(\lambda_1,\dots,\Lambda_L)^\top$ is $L\times R$ and $E$ is the $T\times L$ matrix such that $E_{t\ell}=e_{\ell t}$.
Formally, we first obtain an estimate $\widehat{R}$ of $R$ applying to $X$ one of the estimators of the number of factors available in the literature \citep[see for instance]{bai2002determining,onatski2010determining,ahn2013eigenvalue,bai2019rank,fan2022estimating}. Then, we let the columns of $\widehat{F}/\sqrt{T}$ be the eigenvectors corresponding to the leading $\widehat{R}$ eigenvalues of $XX^\top$. The estimate $\widehat{f}_t\in\R^{\widehat{R}}$ of $f_{t}$ corresponds to the $t^{th}$ colmun of $\widehat{F}^\top$. We note that an alternative estimation procedure for the factors would be to use cross-section averages as in \cite{pesaran2006estimation}. 

Once, we obtained an estimate $\widehat{f}_t$ of the factors, the idea is to leverage the fact that 
$y_{it}=\lambda_{*i}(d_{it})^\top f_t+\tilde u_{it}$ to estimate $\lambda_{*i}$. To simplify estimation, we assume that 
\begin{equation}\label{param}
\lambda_{*ir}(d)=\beta_{i0r}+ \sum_{j=1}^J \beta_{ijr}\varphi_{j}(d),\ d\in\R,\ r=1,\dots,R,
\end{equation}
where $\varphi_j:\R\mapsto \R, j\in[J]$  are $J-1$ known non constant functions of $d$ and $\beta_{irj}\in\R$ are unknown and need to be estimated. For instance, one can pick $J=2$ and $\varphi_1(d)= d,\varphi_2(d)=d^2$, which yields $\lambda_{*ir}(d)=\beta_{i1r}+\beta_{i2r}d+\beta_{i2r}d^2$, allowing for nonlinearity of the effect of $d$ on $\lambda_{*ir}(\cdot)$. We note that $\lambda_{*ir}(\cdot)$ is identified without imposing \eqref{param} and the latter is only useful for estimation. In the spirit of the literature on series estimation, it would be possible to let $J$ grow with the sample size and to assume instead that $\lambda_{i*}(d)$ is only approximated by $ \sum_{j=1}^J \beta_{jr}\varphi_{j}(d)$. We do not do so in the present paper in order to simplify the exposition and the methodology. 

We obtain 
$$y_{it}= \sum_{r=1}^R \beta_{i0r} f_{tr} +\sum_{j=1}^J\sum_{r=1}^R \beta_{ijr}\varphi_{j}(d_{it}) f_t+ \tilde u_{it},$$
where $\tilde u_{it}=\tilde u_{it}(d_{it})$.
Thanks to this, we could estimate $\beta_i=(\beta_{i11},\dots,\beta_{iR1},\dots,\beta_{i1J},\dots,\beta_{iRJ})^\top$ by regressing linearly $y_{it}$ on $f_t$ and $\varphi_j(d_{it})f_t,\ j\in[J]$. Such an approach would only be valid under the exogeneity assumption $\E_t[f_t\tilde u_{it}]= \E_t[\varphi_{j}(d_{it})f_{t}\tilde u_{it}]=0,\ j\in[J]$. This assumption may not always be credible. To weaken it, we include observed controls $c_{it}\in \R^{d_{c}}$, such that $\tilde{u}_{it} =\alpha_i^\top c_{it} +u_{it}$, with $\E_t[f_t u_{it}]= \E_t[\varphi_{j}(d_{it})f_{t} u_{it}]=\E_t[c_{it}u_{it}]=0,\ j\in[J]$ and $\alpha_i\in\R^{d_c}$. The controls $c_{it}$ can belong to $X$ as is standard in panel data models with interactive fixed effects where the variables used to estimate the factors also act as controls \citep{pesaran2006estimation,greenaway2012asymptotic}. The second step regression model can then be written 
\begin{align*} y_{it}&= \sum_{r=1}^R \beta_{i0r} f_{tr}  +\sum_{j=1}^J\sum_{r=1}^R \beta_{ijr}\varphi_{j}(d_{it}) f_t+ \alpha_i^\top c_{it} +u_{it}\\
&= \gamma_i^\top w_{it} +u_{it},
\end{align*}
where $\gamma_i=(\beta_{i}^\top,\alpha_i^\top)^\top$ and $$w_{it}=(1_{R}^\top, \varphi_1(d_{it})f_t^\top, \dots,\varphi_J(d_{it})f_t^\top,c_{it}^\top)$$ is $((J+1)R+d_c)\times 1$.
A natural estimator $\widehat{\gamma}_i$ of $\gamma_i$ is then the linear regression of $y_{it}$ on 
\begin{equation}\label{defhatw} \widehat{w}_{it}=(\widehat{f}_{t}^\top,\varphi_1(d_{it})\widehat{f}_t^\top,\dots,\varphi_J(d_{it}) \widehat{f}_t^\top, c_{it}^\top)^\top\end{equation} a $((J+1)\widehat{R}+d_c)$-dimensional vector. Formally,
\begin{equation}\label{defhatg}\widehat{\gamma}_i= \left(\sum_{t=1}^T \widehat{w}_{it}\widehat{w}_{it}^\top \right)^{-1}\sum_{t=1}^T\widehat{w}_{it} y_{it}.\end{equation}
  Note that, by \eqref{AME2} and \eqref{param}, we have $\Delta_i=\gamma_i^\top\E_t[z_{it}]$, where $$z_{it}=(\varphi_1'(d_{it})f_t^\top,\dots,\varphi_J'(d_{it}) f_t^\top,0_{d_c}^\top)^\top$$ is $((J+1)R+d_c)\times 1$ and $\varphi_j'(\cdot)$ is the derivative of $\varphi_j(\cdot)$. By plug-in the final estimator of $\Delta_i$ is 
\begin{equation}\label{defhatdi} \widehat{\Delta}_i =\widehat{\gamma}_i^\top\left(\frac1T\sum_{t=1}^T \widehat{z}_{it}\right),\end{equation}
where \begin{equation}\label{defhatz} \widehat{z}_{it}=(\varphi_1'(d_{it})\widehat{f}_t^\top,\dots,\varphi_J'(d_{it}) \widehat{f}_t^\top,0_{d_c})^\top\end{equation} is a $(J+1)\widehat{R}+d_c$-dimensional vector. When $N$ is large, we can also estimate $\Delta_t$ by \begin{equation}\label{defhatdt} \widehat{\Delta}_t=\widehat{\gamma}^\top\left( \frac1N\sum_{i=1}^N \widehat{z}_{it} \right),\end{equation}
where $$\widehat{\gamma}=\frac1N \sum_{i=1}^N \widehat{\gamma}_i$$ is an estimator of $\gamma=\E[\gamma_i].$
To conclude, we estimate $\Delta$ by the sample average of the $\widehat{\Delta}_i$, that is 
\begin{equation}\label{defhatd} \widehat{\Delta}=\frac{1}{N} \sum_{i=1}^N\widehat{\Delta}_i.\end{equation}
Algorithm \ref{algo} summarizes the estimation procedure.
\begin{algorithm}
\singlerule\vspace{0.1cm}
\begin{algorithmic}
  \STATE \textbf{1.} Estimate $\widehat{R}$ by one of the available estimators of the number of factors.
  \STATE \textbf{2.} Let the columns of $\widehat{F}/\sqrt{T}$ be the eigenvectors corresponding to the leading $\widehat{R}$ eigenvalues of $XX^\top$.
   \STATE \textbf{3.} Compute $\widehat{w}_{it}$ and $\widehat{\gamma}_i$ according to \eqref{defhatw} and \eqref{defhatg}.
   \STATE \textbf{4.} Compute $\widehat{z}_{it}$ and $\widehat{\Delta}_i$ as in \eqref{defhatz} and \eqref{defhatdi}.
      \STATE \textbf{5.} (Optional) Compute $\widehat{\Delta}_t$ as in \eqref{defhatdt}.
   \STATE \textbf{6.} (Optional) Compute $\widehat{\Delta}$ as in \eqref{defhatd}.
\end{algorithmic}
\doublerule
\caption{Estimation procedure.}
\label{algo}
\end{algorithm}

\section{Theory}\label{sec.th}
We place ourselves in an asymptotic regime where $L,T\to \infty$. When we derive the properties of $\widehat{\Delta}_t$ and$\widehat{\Delta}$ , we additionally assume that $N\to\infty$. The number of factors $R$ is fixed with $T$. It would be possible to let it grow and/or allow for nonstrong factors, see, for instance, \cite{beyhum2019square,beyhum2022factor,freeman2023linear,bai2023approximate}. In this section, as is standard in the literature, we suppose that we know $R$, that is $\widehat{R}=R$ almost surely. Our results would nevertheless stay valid as long as $\P(\widehat{R}=R)\to 1$.

\subsection{Assumptions} 
We make different types of assumptions. First, there are some standard assumptions for factor models on $F,\Lambda$ and $E$. These assumptions correspond to that made in \cite{bai2023approximate} and are therefore not new. These are stated in Section \ref{subsubsec.std}. Additional assumptions, not directly imposed on the factor model consisting in $F,\Lambda$ and $E$ are then outlined in Section \ref{subsubsec.new}
\subsubsection{Standard assumptions on $\bf{F,\Lambda,E}$} \label{subsubsec.std}
We let $e_t=(e_{1t},\dots, e_{Lt})^\top$ and $e_{\ell}=(e_{\ell 1},\dots, e_{\ell T})^\top$
\begin{Assumption}\label{A1}Let $C>0$ not depending on $N,T,L$ and define $\delta_{LT}=\min(\sqrt{L},\sqrt{T})$. The following holds:
\begin{itemize}
\item[(i)] Mean independence: $\E[e_{\ell t}|\lambda_\ell,f_t]=0$ a.s.
\item[(ii)] Weak (cross-sectional and serial) correlation in the errors.
\begin{itemize}
\item[(a)] $\E\left[\frac{1}{\sqrt{L}} \sum_{\ell=1}^L \{e_{\ell t}e_{\ell s} - \E[e_{\ell t}e_{\ell s}]\}\right]$
\item[(b)] For all $\ell$, $\frac{1}{T} \sum_{s,t=1}^T |\E[e_{\ell t}e_{\ell s}] |\le C$. For all $t$, $\frac{1}{L} \sum_{\ell,k=1}^L |\E[e_{\ell t}e_{kt}] |\le C$. 
\item[(c)] For all $t$, $\frac{1}{L\sqrt{T}} \|e_t^\top E^\top\|_2=O_P(\delta_{LT}^{-1})$. For all $i$, $\frac{1}{T\sqrt{L}} \|e_\ell^\top E\|_2=O_P(\delta_{LT}^{-1})$. 
\item[(d)] $\|E\|_{op}^2=O_P(\max(L,T)).$
\end{itemize}
\end{itemize}
\end{Assumption}
\begin{Assumption}\label{A2}
\begin{itemize}
\item[(i)] $\E[||f_t||_2^4] \le C$, $\frac{F^\top F}{T}\xrightarrow[T\to\infty]{\P}\Sigma_F>0$.
\item[(ii)] $\|\lambda_\ell||_2\le C$, $\frac{\Lambda^\top \Lambda}{L}\xrightarrow{}\Sigma_\Lambda>0$.
\item[(iii)] The eigenvalues of $\Sigma_\Lambda\Sigma_F$ are distinct.
\end{itemize}
\end{Assumption}

\begin{Assumption}\label{A3}
For each t, (i) $\E\left[||L^{1/2}\sum_{\ell=1}^L \lambda_\ell  e_{\ell t}||_2^2\right]\le C$, (ii) $\frac{1}{LT} e_tE^\top F=O_P(\delta_{LT}^{-2})$,  for each i, (iii) $\E[\|T^{-1/2}\sum_{t=1}^T f_t e_{\ell t}\|_2^2]\le C$, (iv) $\frac{1}{LT} e_tE\Lambda=O_P(\delta_{LT}^{-2})$, (v) $\Lambda^\top E^\top F=O_P(\sqrt{LT}).$
\end{Assumption} 
We refer the reader to \cite{bai2023approximate} for a discussion of Assumptions \ref{A1} to \ref{A3}.

\subsubsection{New assumptions}\label{subsubsec.new}
 Let $h_{it}=(u_{it}w_{it}^\top, v_{it}^\top)^\top$, with $v_{it}= z_{it}-\E_t[z_{it}]$. The next assumption allows us to show asymptotic normality of $\Delta_i$.
\begin{Assumption}\label{A4} Uniformly in $i\in[N]$, the following holds:
\begin{itemize}
\item[(i)] For all $j\in[J]$, $\varphi_j$ is differentiable.
\item[(ii)] There exists a constant $C>0$ independent of $L,N$ and $T$ such that, for all $j\in[J]$, $|\varphi_j(d_{it})|\le C$ and $|\varphi_j'(d_{it})|\le C$ almost surely.
\item[(iii)] For all $j,j'\in[J]$, \begin{itemize}
\item[(a)] $\sum_{t=1}^T \sum_{\ell =1}^L\varphi_j'(d_{it})  e_{\ell t}\lambda_i=O_P(\sqrt{LT}) $,
\item[(b)] $\sum_{t=1}^T \varphi_j(d_{it})^2=O_P(T)$ and $\sum_{t=1}^T \|c_{it}\|_2^2=O_P(T)$,
\item[(c)] $\sum_{t=1}^T \sum_{\ell=1}^L\varphi_j(d_{it})\varphi_{j'}(d_{it})f_t e_{\ell t}\lambda_i=O_P(\sqrt{LT})$ and $\sum_{t=1}^T \sum_{\ell =1}^Lc_{it} f_t e_{\ell t}\lambda_i=O_P(\sqrt{LT})$ 
\item[(d)] $\sum_{t=1}^T \varphi_j(d_{it})^2\varphi_{j'}(d_{it})^2\|f_t\|_2^2  =O_P(T),$
\item[(e)] $\sum_{t=1}^T \varphi_j(d_{it})^2\varphi_j(d_{it})^2u_{it}^2 =O_P(T),$
\item[(f)] $\sum_{t=1}^T \sum_{\ell=1}^L \varphi_j(d_{it})^2\varphi_j(d_{it})u_{it}e_{\ell t}\lambda_i =O_P(\sqrt{LT}).$
\end{itemize}
\item[(iv)] $\sqrt{T}/L\to 0$. 
\item[(v)] \begin{itemize}
\item[(a)] $\frac1T \sum_{t=1}^T w_{it} w_{it}^\top\xrightarrow[T,L\to\infty]{\P} \Sigma_{w_iw_i}>0,$, 
\item[(b)]  $\frac1T \sum_{t=1}^T h_{it}\xrightarrow[T,L\to\infty]{d} \mathcal{N}\left(0,\Sigma_{h_ih_i}\right),$
\item[(c)] $\frac1T\sum_{t=1}^T z_{it}\xrightarrow[T,L\to\infty]{\P} \E_t[z_{it}].$
\end{itemize}
\end{itemize}
\end{Assumption}
Assumption \ref{A4} (iii) contains mild regularity conditions. The rate condition in Assumption \ref{A4} (iv) is the same as the one used in the literature on factor-augmented regression to show that the error made in estimating the factors is asymptotically negligible \citep{bai2006confidence}. Assumption \ref{A4} (v) states that certain law of large numbers and central limit theorems hold.
Let $$q_{\ell t}=\E_i[\gamma_i^\top b_{i\ell t}] ,$$
where 
$$ b_{i\ell t}= ( 0_R^\top, \varphi_{1}'(d_{it}) g_{\ell t}^\top ,\dots, \varphi_{J}'(d_{it}) g_{\ell t}^\top )\ \text{with } g_{\ell t}=\Sigma_\Lambda^{-1} \lambda_\ell e_{\ell t}.$$
We leverage the next assumption only to derive the asymptotic distribution of $\widehat{\Delta}_t$. 
\begin{Assumption}\label{A5} The following holds:
\begin{itemize}
\item[(i)] For all  $j\in[J]$, \begin{itemize}
\item[(a)] $\frac{1}{NT}\sum_{t=1}^T \sum_{i=1}^N \E_i[z_{it}]^\top\Psi^\top(\Psi^\top \Sigma_{w_iw_i}\Psi)^\top\Psi^\top w_{it} u_{it} =O_P\left(\frac{1}{\sqrt{NT}}\right) $,
\item[(b)] $\frac{1}{\sqrt{NL}}  \sum_{i=1}^N \sum_{\ell=1}^L\gamma_i^\top  b_{i\ell t}-\E_i[\gamma_i^\top b_{i\ell t}]  =O_P(1),$
\end{itemize}
\item[(ii)] $\sqrt{N}/\min(T,L)\to 0$. 
\item[(iii)] \begin{itemize}
\item[(a)] $\frac{1}{\sqrt{N}}\sum_{i=1}^N \gamma_i^\top z_{it} -\Delta_t \xrightarrow[N,T\to\infty]{d} \mathcal{N}(0,\text{var}_i(\gamma_i^\top z_{it}))$, 
\item[(b)] $\frac{1}{\sqrt{L}} \sum_{\ell=1}^L q_{\ell t}\xrightarrow[N,T\to\infty]{d}  \mathcal{N}(0,\sigma^2_{q_t})$,
\item[(c)] $\{\gamma_i^\top z_{it}\}_i$ and $\{q_{\ell t}\}_\ell$ are independent.
\end{itemize}
\end{itemize}
\end{Assumption}
Assumption \ref{A6} (ii) can only be satisfied when both $T$ and $L$ go to infinity. If $X$ consists of $K$ variables for each unit, we have $L=NK$ and the condition simplifies to $\sqrt{N}/T\to 0$.
Let $$m_t=(0_R^\top,\E_i[\varphi_{1}'(d_{it}) ]f_t^\top-\E[ \varphi_1'(d_{it})f_t]^\top,\dots,\E_i[\varphi_{J}'(d_{it}) ]f_t^\top-\E[ \varphi_J'(d_{it})f_t]^\top,0_{d_c}^\top)^\top.$$ The next assumption is only used to establish asymptotic normality of $\widehat{\Delta}$.
\begin{Assumption}\label{A6} The following holds:
\begin{itemize}
\item[(i)] For all  $j\in[J]$, \begin{itemize}
\item[(a)] $\frac{1}{NT}\sum_{t=1}^T \sum_{i=1}^N \E_i[z_{it}]^\top\Psi^\top(\Psi^\top \Sigma_{w_iw_i}\Psi)^\top\Psi^\top w_{it} u_{it} =O_P\left(\frac{1}{\sqrt{NT}}\right) $,
\item[(b)] $\frac{1}{\sqrt{NT}}  \sum_{i=1}^N \sum_{t=1}^T(\gamma_i-\gamma)^\top  v_{it}  =O_P(1);$
\item[(c)] $\frac{1}{\sqrt{NT}}\sum_{i=1}^N  \sum_{t=1}^T(\varphi_{j}'(d_{it})-\E_i[\varphi_j'(d_{it})])f_t=O_P(1).$
\end{itemize}
\item[(ii)] $\sqrt{N}/\min(T,L)\to 0$. 
\item[(iii)] \begin{itemize}
\item[(a)] $\frac{1}{\sqrt{N}}\sum_{i=1}^N \Delta_i -\Delta\xrightarrow[N,T\to\infty]{d} \mathcal{N}(0,\text{var}(\Delta_i))$, 
\item[(b)] $\frac{1}{\sqrt{T}} \sum_{t=1}^T m_{t}\xrightarrow[N,T\to\infty]{d}  \mathcal{N}(0,\Sigma_{mm})$,
\item[(c)] $\{\Delta_i\}_i$ and $\{m_t\}_t$ are independent.
\end{itemize}
\end{itemize}
\end{Assumption}

\subsection{Asymptotic distributions}
To state the asymptotic distribution of our estimators, we need more notation. Let $$D=\text{diag}((\sqrt{\sigma_1(\Sigma_\Lambda\Sigma_F)},\dots,\sqrt{\sigma_R(\Sigma_\Lambda\Sigma_F)})^\top)$$ be the diagonal matrix consisting in the ordered square-root of the eigenvalues of $\Sigma_\Lambda\Sigma_F$. Moreover, we introduce $Q= D \Upsilon^\top \Sigma_{\Lambda}^{-1/2}$, where $\Upsilon=(u_1(\Sigma_F^{1/2} \Sigma_\Lambda \Sigma_F^{1/2}),\dots, u_R(\Sigma_F^{1/2} \Sigma_\Lambda \Sigma_F^{1/2}))$ consists in the eigenvectors of the matrix $\Sigma_F^{1/2} \Sigma_\Lambda \Sigma_F^{1/2}$. We also let $\Psi$ be the $((J+1)R+d_c)\times ((J+1)R+d_c)$ matrix defined by 
$$\Psi =\left(\begin{array}{ccccc} Q^{-1} & 0_{R,R} & \dots & 0_{R,R} & 0_{d_c,d_c}\\ 
0_{R,R} & Q^{-1}& \dots & 0_{R,R} & 0_{d_c,d_c}\\
\vdots & \dots &\ddots &\dots &\vdots\\
0_{R,R}& 0_{R,R} & \dots & Q^{-1} & 0_{d_c,d_c}\\
0_{d_c,R} & 0_{d_c,R} & \dots & 0_{d_c,R} & I_{d_c}
\end{array} \right).$$

Next, we define the  $R\times R$ matrix $\widehat{D}$ whose $r^{th}$ diagonal is the $r^{th}$ singular value of $X/\sqrt{TL}$ and  introduce $\widehat{H}=\left(\frac{\Lambda^\top\Lambda}{N} \right)\left(\frac{F^\top\widehat{F}}{T} \right)\widehat{D}^{-2}.$ We also denote by $\widehat{\Psi}$ the $((J+1)R+d_c)\times ((J+1)R+d_c)$ matrix defined as $\Psi$ but replacing the $Q^{-1}$ matrices by $\widehat{H}$. Finally, define the vector
$\omega_i= \left( \E_t[z_{it}]^\top\Psi^\top(\Psi^\top \Sigma_{w_iw_i}\Psi)^\top,  (\Psi^{-1}\gamma_i)^\top \right)^\top.$   

We show in the Appendix that, under our assumptions, $\widehat{\Psi}\xrightarrow[T,L\to\infty]{\P}\Psi$ and that $\widehat{w}_{it}$, $\widehat{z}_{it}$ and $\widehat{\gamma}_{i}$ estimate consistently $\widehat{\Psi}^\top w_{it}$, $\widehat{\Psi}^\top z_{it}$ and $\widehat{\Psi}^{-1} \gamma_i$, respectively. This allows to derive the asymptotic distributions of our estimators.

The following theorems give the asymptotic distributions of our estimators. First, considering an asymptotic regime with $L,T\to \infty$, we have the following result.

\begin{Theorem} \label{AN}Under Assumptions \ref{A1}-\ref{A4}, we have $$\sqrt{T}(\widehat{\Delta}_i-\Delta_i)=\omega_i^\top \Psi^\top\left(\frac{1}{\sqrt{T}}\sum_{t=1}^Th_{it} \right)+o_P(1), $$
so that 
$\sqrt{T}(\widehat{\Delta}_i-\Delta_i)\xrightarrow[T,L\to\infty]{d} \mathcal{N}\left( 0,\sigma_i^2\right),$
where $\sigma_i^2=\omega_i^\top \Psi^\top \Sigma_{h_ih_i}\Psi  \omega_i$.
\end{Theorem} 
Next, we consider an asymptotic regime with $N,T,L$ all jointly going to infinity.
\begin{Theorem} \label{ANMG}Under Assumptions \ref{A1}-\ref{A4}, if $N/T\to c<\infty $, the following holds 
\begin{itemize}
\item[(i)] If in addition Assumption \ref{A5} is satisfied, we have 
 $$\sqrt{N}(\widehat{\Delta}_t-\Delta_t)=\sqrt{\frac{N}{L}}\left(\frac{1}{\sqrt{L}} \sum_{\ell=1}^L q_{\ell t} \right)+\frac{1}{\sqrt{N}}\sum_{i=1}^N \gamma_i^\top z_{it} -\Delta_t+o_P(1), $$
so that 
$\sqrt{N}(\widehat{\Delta}_t-\Delta_t)\xrightarrow[N,T\to \infty]{d} \mathcal{N}\left( 0,\sigma_t^2\right),$
where $\sigma_t^2=c  \sigma^2_{q_t} + \text{var}_i(\gamma_i^\top z_{it})$.
\item[(ii)] If in addition Assumption \ref{A6} is satisfied, we have 
 $$\sqrt{N}(\widehat{\Delta}-\Delta)=\sqrt{\frac{N}{T}} \gamma^\top\left(\frac{1}{\sqrt{T}}\sum_{t=1}^T m_{t} \right)+\frac{1}{\sqrt{N}}\sum_{i=1}^N \Delta_i-\Delta+o_P(1), $$
so that 
$\sqrt{N}(\widehat{\Delta}-\Delta)\xrightarrow[N,T\to \infty]{d} \mathcal{N}\left( 0,\sigma^2\right),$
where $\sigma^2=c \gamma^\top \Sigma_{mm} \gamma+ \text{var}(\Delta_i)$.
\end{itemize}
\end{Theorem} 
In Theorem \ref{ANMG}, we assume that $N/T\to c<\infty$ just to have a well defined limiting variance. Note that $c$ can be equal to $0$, so that $N$ can be negligible with respect to $T$.
\subsection{Estimation of the variance}\label{sec.estsig}
Let us now discuss estimation of $\sigma_i,\sigma_t$ and $\sigma$. This is crucial to use the theorems for inference.\\

\noindent \textbf{Estimation of $\bf{\sigma_i}$.}  We can estimate $\omega_i$ by 
$$\widehat{\omega}_i=\left( \left(\frac1T\sum_{t=1}^T \widehat{z}_{it}\right)^\top\left(\frac1T\sum_{t=1}^T \widehat{w}_{it}\widehat{w}_{it}^\top\right)^\top, \widehat{\gamma}_i^\top \right)^\top.$$
It remains to estimate $ \Sigma_{h_i}.$ First, we can define $\widehat{h}_{it}=(\widehat{u}_{it}\widehat{w}_{it}^\top, \widehat{v}_{it}^\top)^\top,$ where $\widehat{u}_{it}=y_{it}-\widehat{\gamma}_i^\top w_{it}$. There are different potential estimators of $\Sigma_{h_ih_i}$. If the $h_{it}$ are not serially correlated, one can use the ``standard'' heteroscedasticity consistent (HC) estimator:
$$ \widehat{\Sigma}^{HC}_{h_ih_i}=\frac{1}{T} \sum_{t=1}^T \widehat{h}_{it} \widehat{h}_{it}^\top.$$ 
However, in case of autocorrelation in $h_{it}$ (which may be caused by serial correlation of the factors), one should use an  heteroscedasticity and autocorrelation consistent (HAC) estimator à la \cite{newey1987a}:
$$ \widehat{\Sigma}^{HAC}_{h_ih_i}=\sum_{j=1}^Tk\left(\frac{j}{b}\right)\widehat{\Gamma}_{h_ij},$$
where $\widehat{\Gamma}_{h_ij}= \frac{1}{T}   \sum_{t=1}^{T-j} \widehat{h}_{it+j} \widehat{h}_{it}^\top$, $k$ is a kernel function and $b>0$ is a bandwidth. The final estimate of $\sigma_i^2$ is given by 
$$ \widehat{\sigma}_i^2= \widehat{\omega}_i^\top \widehat{\Sigma}_{h_ih_i} \widehat{\omega}_i,$$ where $\widehat{\Sigma}_{h_ih_i} $ is either $\widehat{\Sigma}^{HC}_{h_ih_i}$ or $\widehat{\Sigma}^{HAC}_{h_ih_i} $.\footnote{An alternative approach is to use fixed-b critical values as in \cite{lazarus2018har}. This would require working out the asymptotic distribution of the test statistic under fixed-b asymptotic, which seems challenging in the present context.}\\

\noindent \textbf{Estimation of $\bf{\sigma_t}$.} Our estimator of $\Lambda$ is the $L\times \widehat{R}$ matrix $\widehat{\Lambda} =(\widehat{\lambda}_1,\dots, \widehat{\lambda}_L)^\top=X^\top \widehat{F}/T$. We also let $\widehat{e}_{\ell t}= x_{\ell t}-\widehat{\lambda}_\ell^\top \widehat{f}_t$ be our estimator of $e_{\ell t}$. These are the standard PCA-based estimators. The estimators of $g_{\ell t}$, $b_{i\ell t}$ and $q_{\ell t}$ are then respectively, \begin{align*}\widehat{g}_{\ell t} &=\left(\frac{\widehat{\Lambda}^\top\widehat{\Lambda}}{N}\right)^{-1}\widehat{\lambda}_\ell \widehat{e}_{\ell t};\\
\widehat{b}_{i\ell t} &= ( 0_R^\top, \varphi_{1}'(d_{it}) \widehat{g}_{\ell t}^\top ,\dots, \varphi_{J}'(d_{it}) \widehat{g}_{\ell t}^\top );\\
                        \widehat{q}_{\ell t}&=\frac1N\sum_{i=1}^N\widehat{\gamma}_i^\top \widehat{b}_{i\ell t} .
                        \end{align*}
                  Let us denote by $$\widehat{\sigma}_{q_{t}}^2= \frac{1}{L}\sum_{\ell =1}^L \widehat{q}_{\ell t} \widehat{q}_{\ell t}^\top$$
                  the estimator of $\sigma_{q_t}^2$.\footnote{This estimator will be consistent when $\{q_{\ell t}\}_{\ell}$ are uncorrelated across $\ell$. This will happen if the $\{e_{\ell t}\}_{\ell}$ are uncorrelated across $\ell$ and independent of the loadings. This is an assumption that could be considered natural in factor models, and the results should not be too sensitive to mild violations of this assumption. To avoid such an assumption, one could use a cross-section consistent estimator as in \cite{bai2006confidence}.} We can estimate $\text{var}_i(\gamma_i^\top z_{it})$ by $$\widehat{\text{var}}_i(\gamma_i^\top z_{it})=\frac1N\sum_{i=1}^N \left(\widehat{\gamma}_i^\top \widehat{z}_{it}- \widehat{\Delta}_t\right)^2.$$ The final estimator of $\sigma_{t}^2$ is then $$\widehat{\sigma}_t^2= \left(\frac{N}{L}\right) \widehat{\sigma}_{q_{t}}^2+ \widehat{\text{var}}_i(\gamma_i^\top z_{it}).$$
                  
\noindent \textbf{Estimation of $\bf{\sigma}$.}  The variable $ \Psi^\top m_t$ can be estimated by 
$$\widehat{m}_t= \left(0_R^\top,\widehat{m}_{t1}^\top,\dots,\widehat{m}_{tJ}^\top,0_{d_c}\right)^\top,$$
where $$\widehat{m}_{tj}=\left(\sum_{i=1}^N \varphi_{j}'(d_{it}) \right)\widehat{f}_t^\top-\left(\frac{1}{NT} \sum_{i=1}^N\sum_{t=1}^T \varphi_j'(d_{it})\widehat{f}_t\right)^\top.$$ A natural estimator of $\sigma^2$ is 
$$\widehat{\sigma}^2=\frac{N}{T}\widehat{\gamma}^\top \widehat{\Sigma}_{mm}\widehat{\gamma}+ \widehat{\text{var}}(\Delta_i),$$
where $\widehat{\gamma}=N^{-1}\sum_{i=1}^N \widehat{\gamma}_i$ estimates $\Psi^{-1}\gamma$, $\widehat{\text{var}}(\Delta_i)=N^{-1}\sum_{i=1}^N (\widehat{\Delta}_i-\widehat{\Delta})^2$ and $ \widehat{\Sigma}_{mm}$ is either
$$\widehat{\Sigma}^{HC}_{mm} = \frac1T\sum_{t=1}^T \widehat{m}_t\widehat{m}_t^\top$$
or, in case of serial correlation, 
$$ \widehat{\Sigma}^{HAC}_{mm}=\sum_{j=1}^Tk\left(\frac{j}{b}\right)\widehat{\Gamma}_{mj},$$
where $\widehat{\Gamma}_{mj}= \frac{1}{T}   \sum_{t=1}^{T-j} \widehat{m}_{t+j} \widehat{m}_{t}^\top$, $k$ is a kernel function and $b>0$ is a bandwidth.

\section{Extensions} \label{sec.ext} In this section, we discuss extensions of our approach to other target parameters and IV estimation. This showcases the flexibility of our modelling strategy.\\

\noindent \textbf{Average counterfactuals.} Let us now discuss other interesting parameters. First, one may want to know what would be the average outcome if the treatment was fixed at $d\in\R$. The following parameters answer this question:
$$\varphi_i(d)=\E_t[y_{it}(d)],\ \varphi_t(d)=\E_i[y_{it}(d)],\ \varphi(d)=\E[y_{it}(d)].$$
The mappings $\varphi_i(\cdot)$, $\varphi_t(\cdot)$ and $\varphi(\cdot)$ are functional parameters and they can be estimated as follows:
$$\widehat{\varphi}_i(d)=\widehat{\gamma}_i^\top\left(\frac1T\sum_{t=1}^T \widehat{w}_{it}(d)\right),\,\ \widehat{\varphi}_t(d)=\widehat{\gamma}^\top\left( \frac1N\sum_{i=1}^N \widehat{w}_{it}(d) \right),\ \varphi(d)=\frac1N\sum_{i=1}^N \widehat{\varphi}_i(d),$$
where $$ \widehat{w}_{it}(d)=(\widehat{f}_{t}^\top,\varphi_1(d)\widehat{f}_t^\top,\dots,\varphi_J(d) \widehat{f}_t^\top, 0_{d_c}^\top)^\top.$$
The mappings $\varphi_i(\cdot)$, $\varphi_t(\cdot)$ and $\varphi(\cdot)$  can also be differentiated to obtain average marginal effects at fixed d and also integrated (possibly after differentiation) with respect to some relevant distibution of $d$.\\

\noindent \textbf{IV estimation.} Suppose now that $d_{it}$ is endogenous even after including the controls, that is $\E_t[\varphi_{j}(d_{it})f_{t} u_{it}]\ne 0$ for some $j\in[J]$. A remedy to this problem is to rely on an IV $s_{it}\in \R$. Let us define $\widehat{r}_{it} = (\widehat{f}_{t}^\top,\varphi_1(s_{it})\widehat{f}_t^\top,\dots,\varphi_J(s_{it}) \widehat{f}_t^\top ,c_{it}^\top)^\top.$ To estimate $\gamma_i$ one can then use the IV estimator
$$\widehat{\gamma}_i^{IV}=\left(\sum_{t=1}^T \widehat{r}_{it}\widehat{w}_{it}^\top \right)^{-1}\sum_{t=1}^T\widehat{r}_{it} y_{it}.$$
The estimators of the other parameters of interest can then be obtained by plug-in. This approach relies on the exogeneity assumption: $\E_t[f_t u_{it}]= \E_t[\varphi_{j}(s_{it})f_{t} u_{it}]=\E_t[c_{it}u_{it}]=0$.

\section{Simulations}\label{sec.sim}
In this section, we provide a Monte Carlo study which sheds light on the finite sample performance of our proposed inference procedures. We start by analyzing $\widehat{\Delta}_i$ in Section \ref{subsec.sing} before studying $\widehat{\Delta}_t$ and $\widehat{\Delta}$ in Section \ref{subsec.panel}.
\subsection{Single unit }\label{subsec.sing} 
To analyze $\widehat{\Delta}_i$, we restrict ourselves to a sample with a single unit $i=1$, that is $N=1$. We consider $T\in\{50,100, 200\}$ and $L\in\{50,100,200\}$. We generate samples with $R=2$ factors.  The loadings are i.i.d. such that $\lambda_{\ell r}\sim\mathcal{U}[-1,1], \ell\in[L], r\in[R]$. The factors are generated as $f_{tr}=0.5+\rho_f^r (f_{t-1r}-0.5)+ \tilde f_{tr}$ for $t=2,\dots, T$ and $r=1,2$, where $\tilde f_{tr}$ are i.i.d. $\mathcal{N}\left(0,I_R\left(1-\rho_f^{2r}\right)\right)$. The stationary distribution of $f_{tr}$ is $\mathcal{N}(0.5,1)$, we initialize $f_{0r}$ as such. The quantity $\rho_{f}$ controls the level of serial correlation and we let $\rho_f\in \{0,0.5\}$. The idiosyncratic components $\{e_{\ell t}\}$ are i.i.d. $\mathcal{N}(0,1)$.  We have two controls, $c_{1t} = (x_{1t},x_{2t})$. We let 
$\lambda_{*1r}(d)=0.5+\sum_{j=1}^J 0.5\times d^{j},$ where $J\in\{1,2\}$. The treatment is such that $d_{1t}=f_{t1}+0.5\times e_{1t}+0.5\times e_{2t}+\epsilon_t$, where $\epsilon_t\sim\mathcal{N}(0,1)$. The outcome is generated as
$$y_{1t} =\lambda_{*1r}(d_{1t})^\top f_t-0.5\times c_{1t}- 0.5\times c_{2t} + u_{1t},$$
where $u_{1t}$ is i.i.d. $\mathcal{N}(0,1)$. We estimate $\Delta_i$ for $i=1$,  which is equal to $0.5$ when $J=1$ and $2$ when $J=2$. We use the growth ratio estimator of \cite{ahn2013eigenvalue} to estimate the number of factors. Following Section \ref{sec.estsig}, we consider three possible estimators of the covariance matrix $\Sigma_{h_ih_i}$. First, there is the standard estimator $\widehat{\Sigma}^{HC}_{h_ih_i}$, second, there is a HAC estimator $ \widehat{\Sigma}^{HAC}_{h_ih_i}$ with quadratic spectral kernel and third a HAC estimator  $ \widehat{\Sigma}^{HAC}_{h_ih_i}$ with Parzen kernel. For the HAC estimators, we pick the bandwidth equal to $1.3T^{1/2}$, which corresponds to the recommendation of \cite{lazarus2018har}.\footnote{The choice in \cite{lazarus2018har} is for linear regression and fixed-b critical values, that is a different context than ours.} For each estimator of the covariance matrix, we then construct 95\% confidence intervals based on the Gaussian approximation of Theorem \ref{AN}.
 
 Tables \ref{tab.singJ1} and \ref{tab.singJ2} present the results for $J=1$ and $J=2$, respectively. These are averages over 8,000 replications. We report the bias, variance, mean-squared error (MSE) of our estimator along with the average radius of $95\%$ confidence intervals built using the three different estimators of the covariance matrix and the Gaussian approximation of Theorem \ref{AN}. We see that our estimator has low bias, variance and MSE, although they worsen when $J$ or the level of autocorrelation in the factors increase. The coverage of the confidence intervals is close to nominal when the factors are not serially correlated $\rho_f=0$. In this case, the differences in radius and coverage between the three types of confidence intervals is minimal. When $\rho_f=0.5$ the coverage worsens for all three types of CI, but the decrease is much steeper for the standard CI not taking into account time series dependence. This suggests than one should indeed use HAC estimators of the covariance matrix in practice, since they greatly improve the results when there is autocorrelation at little cost in the absence of the latter. The CIs based on the Quadratic Spectral and the Parzen Kernel seem to have similar performance. 
\afterpage{
\clearpage
  \thispagestyle{empty}
  \begin{landscape}
\begin{table}\setlength\extrarowheight{-5pt}
\begin{center}
\begin{tabular}{|ll|ccccccccc|}
\hline 
$T$ & $L$ & Bias & Var & MSE & Av. R. CI & Cov. CI &\begin{tabular}{c} Av. R. CI \\ QS ker. \end{tabular}& \begin{tabular}{c}  Cov. CI \\QS ker. \end{tabular}&  \begin{tabular}{c}Av. R. CI\\ Parzen ker.\end{tabular} &   \begin{tabular}{c}Cov. CI\\ Parzen ker.\end{tabular}\\
\hline 
\multicolumn{11}{|c|}{Design 1: $J=1$,\ $\rho_f=0$}\\
\hline
50 & 50 & -0.0171 & 0.0139 & 0.0142 & 0.2194 & 0.92 & 0.21 & 0.89 & 0.21 & 0.90\\
50 & 100 & -0.0064 & 0.0138 & 0.0139 & 0.2179 & 0.93 & 0.21 & 0.90 & 0.21 & 0.91\\
50 & 200 & -0.0051 & 0.0131 & 0.0131 & 0.2162 & 0.93 & 0.20 & 0.91 & 0.21 & 0.92\\
100 & 50 & -0.0187 & 0.0069 & 0.0073 & 0.1547 & 0.92 & 0.15 & 0.90 & 0.15 & 0.90\\
100 & 100 & -0.0102 & 0.0065 & 0.0066 & 0.1537 & 0.93 & 0.15 & 0.91 & 0.15 & 0.92\\
100 & 200 & -0.0049 & 0.0066 & 0.0067 & 0.1526 & 0.93 & 0.15 & 0.92 & 0.15 & 0.92\\
200 & 50 & -0.0182 & 0.0034 & 0.0037 & 0.1098 & 0.92 & 0.11 & 0.91 & 0.11 & 0.92\\
200 & 100 & -0.0102 & 0.0032 & 0.0033 & 0.1085 & 0.93 & 0.11 & 0.92 & 0.11 & 0.92\\
200 & 200 & -0.0039 & 0.0032 & 0.0032 & 0.1080 & 0.94 & 0.10 & 0.93 & 0.11 & 0.93\\
\hline
\multicolumn{11}{|c|}{Design 2: $J=1$,\ $\rho_f=0.5$}\\
\hline
50 & 50 & -0.0171 & 0.0265 & 0.0268 & 0.2198 & 0.81 & 0.24 & 0.84 & 0.25 & 0.85\\
50 & 100 & -0.0068 & 0.0270 & 0.0270 & 0.2178 & 0.81 & 0.24 & 0.84 & 0.25 & 0.85\\
50 & 200 & -0.0050 & 0.0263 & 0.0263 & 0.2160 & 0.81 & 0.24 & 0.84 & 0.24 & 0.85\\
100 & 50 & -0.0197 & 0.0135 & 0.0139 & 0.1545 & 0.80 & 0.18 & 0.84 & 0.18 & 0.85\\
100 & 100 & -0.0106 & 0.0133 & 0.0134 & 0.1539 & 0.80 & 0.18 & 0.86 & 0.18 & 0.87\\
100 & 200 & -0.0055 & 0.0134 & 0.0134 & 0.1526 & 0.81 & 0.18 & 0.86 & 0.18 & 0.87\\
200 & 50 & -0.0181 & 0.0067 & 0.0070 & 0.1099 & 0.81 & 0.13 & 0.86 & 0.13 & 0.87\\
200 & 100 & -0.0105 & 0.0065 & 0.0066 & 0.1086 & 0.82 & 0.13 & 0.87 & 0.13 & 0.88\\
200 & 200 & -0.0034 & 0.0065 & 0.0065 & 0.1081 & 0.82 & 0.13 & 0.88 & 0.13 & 0.88\\
\hline
\end{tabular}
\begin{tablenotes}
 \item[\hskip -\fontdimen 2 \font]Note: ``Av. R. CI'', ``Av. R. CI QS ker.'' and ``Av. R. CI Parzen ker.'' (respectively ``Cov. CI'', ``Cov. CI QS ker.'' and ``Cov. CI Parzen ker.'') stand for the average radius (respectively, coverage) of the 95\% confidence intervals computed using, respectively, the standard estimator, the HAC estimator with quadratic spectral kernel and the HAC estimator with Parzen kernel of the covariance matrix.
\end{tablenotes}
\end{center}
\caption{Results for the estimator of the AME of a single unit $\widehat{\Delta}_i$ with $J=2$} 
\label{tab.singJ1}
\end{table}
  \end{landscape}
  \clearpage
}

\afterpage{
\clearpage
  \thispagestyle{empty}
  
  \begin{landscape}
\begin{table}\setlength\extrarowheight{-5pt}
\begin{center}
\begin{tabular}{|ll|ccccccccc|}
\hline 
$T$ & $L$ & Bias & Var & MSE & Av. R. CI & Cov. CI &\begin{tabular}{c} Av. R. CI \\ QS ker. \end{tabular}& \begin{tabular}{c}  Cov. CI \\QS ker. \end{tabular}&  \begin{tabular}{c}Av. R. CI\\ Parzen ker.\end{tabular} &   \begin{tabular}{c}Cov. CI\\ Parzen ker.\end{tabular}\\
\hline 
\multicolumn{11}{|c|}{Design 1: $J=2$,\ $\rho_f=0$}\\
\hline
50 & 50 & -0.0249 & 0.3829 & 0.3836 & 1.1406 & 0.91 & 1.07 & 0.89 & 1.10 & 0.90\\
50 & 100 & -0.0069 & 0.3740 & 0.3740 & 1.1318 & 0.92 & 1.07 & 0.90 & 1.09 & 0.91\\
50 & 200 & -0.0037 & 0.3576 & 0.3576 & 1.1250 & 0.92 & 1.06 & 0.90 & 1.09 & 0.91\\
100 & 50 & -0.0403 & 0.1945 & 0.1961 & 0.8209 & 0.92 & 0.78 & 0.90 & 0.80 & 0.91\\
100 & 100 & -0.0112 & 0.1808 & 0.1809 & 0.8131 & 0.93 & 0.78 & 0.92 & 0.79 & 0.92\\
100 & 200 & -0.0088 & 0.1805 & 0.1806 & 0.8036 & 0.93 & 0.77 & 0.91 & 0.78 & 0.92\\
200 & 50 & -0.0343 & 0.0979 & 0.0991 & 0.5879 & 0.93 & 0.57 & 0.92 & 0.58 & 0.92\\
200 & 100 & -0.0225 & 0.0915 & 0.0920 & 0.5771 & 0.93 & 0.56 & 0.93 & 0.57 & 0.93\\
200 & 200 & -0.0015 & 0.0909 & 0.0909 & 0.5750 & 0.94 & 0.56 & 0.93 & 0.56 & 0.93\\
\hline
\multicolumn{11}{|c|}{Design 2: $J=2$,\ $\rho_f=0.5$}\\
\hline
50 & 50 & -0.0292 & 0.5863 & 0.5872 & 1.1278 & 0.84 & 1.16 & 0.84 & 1.19 & 0.85\\
50 & 100 & -0.0089 & 0.5767 & 0.5768 & 1.1207 & 0.85 & 1.16 & 0.84 & 1.18 & 0.85\\
50 & 200 & -0.0049 & 0.5572 & 0.5572 & 1.1114 & 0.85 & 1.15 & 0.85 & 1.17 & 0.86\\
100 & 50 & -0.0459 & 0.2931 & 0.2952 & 0.8146 & 0.85 & 0.86 & 0.86 & 0.88 & 0.87\\
100 & 100 & -0.0097 & 0.2874 & 0.2875 & 0.8092 & 0.86 & 0.86 & 0.87 & 0.88 & 0.88\\
100 & 200 & -0.0097 & 0.2866 & 0.2867 & 0.8000 & 0.85 & 0.85 & 0.86 & 0.86 & 0.87\\
200 & 50 & -0.0328 & 0.1487 & 0.1498 & 0.5869 & 0.86 & 0.63 & 0.88 & 0.64 & 0.89\\
200 & 100 & -0.0251 & 0.1418 & 0.1424 & 0.5750 & 0.86 & 0.62 & 0.88 & 0.63 & 0.89\\
200 & 200 & 0.0008 & 0.1435 & 0.1435 & 0.5731 & 0.87 & 0.62 & 0.89 & 0.63 & 0.90\\
\hline
\end{tabular}
\begin{tablenotes}
 \item[\hskip -\fontdimen 2 \font]Note: ``Av. R. CI'', ``Av. R. CI QS ker.'' and ``Av. R. CI Parzen ker.'' (respectively ``Cov. CI'', ``Cov. CI QS ker.'' and ``Cov. CI Parzen ker.'') stand for the average radius (respectively, coverage) of the 95\% confidence intervals computed using, respectively, the standard estimator, the HAC estimator with quadratic spectral kernel and the HAC estimator with Parzen kernel of the covariance matrix.
\end{tablenotes}
\end{center}
\caption{Results for the estimator of the AME of a single unit $\widehat{\Delta}_i$ with $J=2$} 
\label{tab.singJ2}
\end{table}
  \end{landscape}
  \clearpage
}

\subsection{Large panel} \label{subsec.panel} 
Now, we study $\widehat{\Delta}_t$ and $\widehat{\Delta}$. To do so we consider a panel data with $N\in\{50,100,200\}$ units and $T\in\{50,100, 200\}$ dates. We set $L=2N$, which mimicks the case where the panel $X$ consists in $2$ auxiliary variables (corresponding to $\ell=2i-1$ and $\ell=2i$) for each unit $i$ at each date $t$. We generate the factors $f_t$ and idiosyncratic errors $e_{it}$ as in Section \ref{subsec.sing}. There are two controls $c_{it} = (x_{(2i-1)t},x_{(2i)t})^\top$. We let 
$\lambda_{*ir}(d)=\beta_{0i}+\sum_{j=1}^J\beta_{ji}\times d^{j},$ where $\beta_{ji}=0.5+\mathcal{U}[-0.5,0.5]$ and $J\in\{1,2\}$. The treatment is such that $d_{it}=f_{t1}+0.5\times e_{(2i-1)t}+0.5\times e_{(2i)t}+\epsilon_{it}$, where $\epsilon_{it}$ is i.i.d. $\mathcal{N}(0,1)$. The outcome is generated as
$$y_{it} =\lambda_{*1r}(d_{it})^\top f_t-0.5\times x_{(2i-1)t}- 0.5\times x_{(2i)t} + u_{it},$$
where $u_{it}$ is i.i.d. $\mathcal{N}(0,1)$. 

First, we estimate $\Delta_t$ for $t=1$. It is equal to $0.5(f_{11}+f_{12})$ when $J=1$ and $0.5(f_{11}+f_{12})+2\times 0.5(f_{11}^2+f_{11}f_{12})$ when $J=2$. The results are presented in Tables \ref{tab.panetJ1} and \ref{tab.panetJ2} and are, again, averages over 8,000 replications. The variance $\sigma_{q_t}$ is not affected by autocorrelation in the factors and therefore we only report 95\% confidence intervals built using $\widehat{\sigma}_{q_t}$ and Gaussian approximation. The results confirm that the performance of the estimator is indeed not affected by autocorrelation. The bias, variance and MSE of the estimator is again low and the coverage of the confidence intervals close to nominal.

\begin{table}\setlength\extrarowheight{-5pt}
\begin{center}
\begin{tabular}{|ll|ccccc|}
\hline 
$T$ & $N$ & Bias & Var & MSE & Av. R. CI & Cov. CI \\
\hline 
\multicolumn{7}{|c|}{Design 1: $J=1$,\ $\rho_f=0$}\\
\hline
50 & 50 & -0.0093 & 0.0203 & 0.0204 & 0.2640 & 0.93\\
50 & 100 & -0.0081 & 0.0199 & 0.0200 & 0.2649 & 0.93\\
50 & 200 & -0.0087 & 0.0203 & 0.0204 & 0.2659 & 0.94\\
100 & 50 & -0.0029 & 0.0103 & 0.0103 & 0.1882 & 0.94\\
100 & 100 & -0.0030 & 0.0102 & 0.0102 & 0.1893 & 0.94\\
100 & 200 & -0.0043 & 0.0098 & 0.0098 & 0.1901 & 0.95\\
200 & 50 & 0.0003 & 0.0051 & 0.0051 & 0.1337 & 0.94\\
200 & 100 & -0.0030 & 0.0051 & 0.0051 & 0.1349 & 0.94\\
200 & 200 & -0.0010 & 0.0051 & 0.0051 & 0.1359 & 0.95\\
\hline
\multicolumn{7}{|c|}{Design 2: $J=1$,\ $\rho_f=0.5$}\\
\hline
50 & 50 & -0.0094 & 0.0203 & 0.0204 & 0.2638 & 0.93\\
50 & 100 & -0.0080 & 0.0200 & 0.0200 & 0.2647 & 0.93\\
50 & 200 & -0.0087 & 0.0204 & 0.0204 & 0.2658 & 0.93\\
100 & 50 & -0.0029 & 0.0103 & 0.0103 & 0.1881 & 0.94\\
100 & 100 & -0.0032 & 0.0102 & 0.0102 & 0.1892 & 0.94\\
100 & 200 & -0.0042 & 0.0098 & 0.0099 & 0.1901 & 0.94\\
200 & 50 & 0.0003 & 0.0051 & 0.0051 & 0.1336 & 0.94\\
200 & 100 & -0.0031 & 0.0051 & 0.0051 & 0.1348 & 0.94\\
200 & 200 & -0.0011 & 0.0051 & 0.0051 & 0.1359 & 0.94\\
\hline
\end{tabular}
\begin{tablenotes}
 \item[\hskip -\fontdimen 2 \font]Note: ``Av. R. CI'' (respectively ``Cov. CI'') stand for the average radius (respectively, coverage) of the 95\% confidence intervals.
\end{tablenotes}
\end{center}
\caption{Results for the estimator of the time-specific AME $\widehat{\Delta}_t$ with $J=1$} 
\label{tab.panetJ1}
\end{table}

\begin{table}\setlength\extrarowheight{-5pt}
\begin{center}
\begin{tabular}{|ll|ccccc|}
\hline 
$T$ & $N$ & Bias & Var & MSE & Av. R. CI & Cov. CI \\
\hline 
\multicolumn{7}{|c|}{Design 1: $J=2$,\ $\rho_f=0$}\\
\hline
50 & 50 & -0.0044 & 0.2771 & 0.2771 & 0.8814 & 0.94\\
50 & 100 & -0.0084 & 0.2904 & 0.2905 & 0.8745 & 0.94\\
50 & 200 & -0.0101 & 0.2807 & 0.2808 & 0.8761 & 0.94\\
100 & 50 & -0.0069 & 0.1374 & 0.1374 & 0.6229 & 0.94\\
100 & 100 & -0.0021 & 0.1356 & 0.1356 & 0.6162 & 0.94\\
100 & 200 & -0.0008 & 0.1431 & 0.1431 & 0.6179 & 0.94\\
200 & 50 & -0.0024 & 0.0677 & 0.0677 & 0.4361 & 0.94\\
200 & 100 & -0.0007 & 0.0668 & 0.0668 & 0.4383 & 0.95\\
200 & 200 & -0.0042 & 0.0747 & 0.0747 & 0.4459 & 0.94\\
\hline
\multicolumn{7}{|c|}{Design 2: $J=2$,\ $\rho_f=0.5$}\\
\hline
50 & 50 & -0.0047 & 0.2763 & 0.2763 & 0.8813 & 0.94\\
50 & 100 & -0.0084 & 0.2912 & 0.2913 & 0.8746 & 0.94\\
50 & 200 & -0.0103 & 0.2816 & 0.2817 & 0.8760 & 0.94\\
100 & 50 & -0.0073 & 0.1364 & 0.1364 & 0.6228 & 0.94\\
100 & 100 & -0.0026 & 0.1350 & 0.1350 & 0.6160 & 0.94\\
100 & 200 & -0.0008 & 0.1429 & 0.1429 & 0.6179 & 0.94\\
200 & 50 & -0.0026 & 0.0677 & 0.0677 & 0.4359 & 0.94\\
200 & 100 & -0.0011 & 0.0668 & 0.0668 & 0.4382 & 0.95\\
200 & 200 & -0.0041 & 0.0742 & 0.0742 & 0.4458 & 0.94\\
\hline
\end{tabular}
\begin{tablenotes}
 \item[\hskip -\fontdimen 2 \font]Note: ``Av. R. CI'' (respectively ``Cov. CI'') stand for the average radius (respectively, coverage) of the 95\% confidence intervals.
\end{tablenotes}
\end{center}
\caption{Results for the estimator of the time-specific AME $\widehat{\Delta}_t$ with $J=2$} 
\label{tab.panetJ2}
\end{table}

Then we estimate $\Delta$. In this design, we have $\Delta=0.5$ when $J=1$ and $\Delta=2$ when $J=2$. The results are displayed in Tables \ref{tab.panelJ1} and \ref{tab.panelJ2} and are, again, averages over 8,000 replications. As in Section \ref{subsec.sing}, we find that aucocorrelation or increasing $J$ worsen the results but HAC estimators of the covariance matrix can mitigate the decrease in coverage of the CIs.

\afterpage{
\clearpage
  \thispagestyle{empty}
  \begin{landscape}
\begin{table}\setlength\extrarowheight{-5pt}
\begin{center}
\begin{tabular}{|ll|ccccccccc|}
\hline 
$T$ & $N$ & Bias & Var & MSE & Av. R. CI & Cov. CI &\begin{tabular}{c} Av. R. CI \\ QS ker. \end{tabular}& \begin{tabular}{c}  Cov. CI \\QS ker. \end{tabular}&  \begin{tabular}{c}Av. R. CI\\ Parzen ker.\end{tabular} &   \begin{tabular}{c}Cov. CI\\ Parzen ker.\end{tabular}\\
\hline 
\multicolumn{11}{|c|}{Design 1: $J=1$,\ $\rho_f=0$}\\
\hline
50 & 50 & -0.0066 & 0.0122 & 0.0123 & 0.2101 & 0.94 & 0.20 & 0.92 & 0.20 & 0.93\\
50 & 100 & -0.0066 & 0.0067 & 0.0067 & 0.1594 & 0.94 & 0.15 & 0.93 & 0.16 & 0.93\\
50 & 200 & -0.0081 & 0.0042 & 0.0043 & 0.1257 & 0.94 & 0.12 & 0.93 & 0.12 & 0.93\\
100 & 50 & -0.0038 & 0.0108 & 0.0108 & 0.2019 & 0.94 & 0.19 & 0.92 & 0.20 & 0.93\\
100 & 100 & -0.0048 & 0.0058 & 0.0058 & 0.1491 & 0.94 & 0.14 & 0.93 & 0.15 & 0.94\\
100 & 200 & -0.0051 & 0.0034 & 0.0034 & 0.1127 & 0.94 & 0.11 & 0.94 & 0.11 & 0.94\\
200 & 50 & -0.0009 & 0.0105 & 0.0105 & 0.1977 & 0.94 & 0.19 & 0.92 & 0.19 & 0.93\\
200 & 100 & -0.0016 & 0.0054 & 0.0054 & 0.1435 & 0.95 & 0.14 & 0.93 & 0.14 & 0.94\\
200 & 200 & -0.0014 & 0.0029 & 0.0030 & 0.1056 & 0.95 & 0.10 & 0.94 & 0.10 & 0.94\\
\hline
\multicolumn{11}{|c|}{Design 2: $J=1$,\ $\rho_f=0.5$}\\
\hline
50 & 50 & -0.0067 & 0.0253 & 0.0254 & 0.2083 & 0.81 & 0.24 & 0.85 & 0.24 & 0.86\\
50 & 100 & -0.0060 & 0.0132 & 0.0133 & 0.1589 & 0.83 & 0.18 & 0.88 & 0.18 & 0.88\\
50 & 200 & -0.0079 & 0.0075 & 0.0075 & 0.1256 & 0.85 & 0.14 & 0.89 & 0.14 & 0.89\\
100 & 50 & -0.0038 & 0.0237 & 0.0237 & 0.1996 & 0.80 & 0.23 & 0.84 & 0.23 & 0.85\\
100 & 100 & -0.0050 & 0.0123 & 0.0123 & 0.1484 & 0.82 & 0.17 & 0.87 & 0.17 & 0.88\\
100 & 200 & -0.0055 & 0.0067 & 0.0067 & 0.1126 & 0.83 & 0.13 & 0.88 & 0.13 & 0.89\\
200 & 50 & -0.0007 & 0.0233 & 0.0233 & 0.1949 & 0.79 & 0.22 & 0.83 & 0.23 & 0.84\\
200 & 100 & -0.0014 & 0.0119 & 0.0119 & 0.1428 & 0.81 & 0.17 & 0.86 & 0.17 & 0.87\\
200 & 200 & -0.0012 & 0.0063 & 0.0063 & 0.1053 & 0.82 & 0.13 & 0.88 & 0.13 & 0.88\\
\hline
\end{tabular}
\begin{tablenotes}
 \item[\hskip -\fontdimen 2 \font]Note: ``Av. R. CI'', ``Av. R. CI QS ker.'' and ``Av. R. CI Parzen ker.'' (respectively ``Cov. CI'', ``Cov. CI QS ker.'' and ``Cov. CI Parzen ker.'') stand for the average radius (respectively, coverage) of the 95\% confidence intervals computed using, respectively, the standard estimator, the HAC estimator with quadratic spectral kernel and the HAC estimator with Parzen kernel of the covariance matrix.
\end{tablenotes}
\end{center}
\caption{Results for the estimator of the AME $\widehat{\Delta}$ with $J=1$} 
\label{tab.panelJ1}
\end{table}
  \end{landscape}
  \clearpage
}

\afterpage{
\clearpage
  \thispagestyle{empty}
  
  \begin{landscape}
\begin{table}\setlength\extrarowheight{-5pt}
\begin{center}
\begin{tabular}{|ll|ccccccccc|}
\hline 
$T$ & $N$ & Bias & Var & MSE & Av. R. CI & Cov. CI &\begin{tabular}{c} Av. R. CI \\ QS ker. \end{tabular}& \begin{tabular}{c}  Cov. CI \\QS ker. \end{tabular}&  \begin{tabular}{c}Av. R. CI\\ Parzen ker.\end{tabular} &   \begin{tabular}{c}Cov. CI\\ Parzen ker.\end{tabular}\\
\hline 
\multicolumn{11}{|c|}{Design 1: $J=2$,\ $\rho_f=0$}\\
\hline
50 & 50 & -0.0089 & 0.1872 & 0.1872 & 0.8125 & 0.92 & 0.77 & 0.90 & 0.79 & 0.91\\
50 & 100 & -0.0055 & 0.1006 & 0.1006 & 0.6103 & 0.93 & 0.59 & 0.92 & 0.60 & 0.93\\
50 & 200 & -0.0116 & 0.0583 & 0.0584 & 0.4680 & 0.94 & 0.46 & 0.94 & 0.46 & 0.94\\
100 & 50 & -0.0018 & 0.1670 & 0.1670 & 0.7881 & 0.93 & 0.74 & 0.91 & 0.76 & 0.92\\
100 & 100 & -0.0032 & 0.0902 & 0.0902 & 0.5790 & 0.94 & 0.56 & 0.92 & 0.57 & 0.93\\
100 & 200 & -0.0060 & 0.0506 & 0.0507 & 0.4305 & 0.94 & 0.42 & 0.93 & 0.42 & 0.93\\
200 & 50 & 0.0015 & 0.1676 & 0.1676 & 0.7750 & 0.93 & 0.73 & 0.91 & 0.75 & 0.91\\
200 & 100 & -0.0003 & 0.0836 & 0.0836 & 0.5618 & 0.94 & 0.54 & 0.92 & 0.55 & 0.93\\
200 & 200 & -0.0048 & 0.0438 & 0.0438 & 0.4093 & 0.94 & 0.40 & 0.93 & 0.40 & 0.94\\
\hline
\multicolumn{11}{|c|}{Design 2: $J=2$,\ $\rho_f=0.5$}\\
\hline
50 & 50 & -0.0090 & 0.3858 & 0.3859 & 0.7969 & 0.78 & 0.90 & 0.81 & 0.91 & 0.82\\
50 & 100 & -0.0045 & 0.2051 & 0.2051 & 0.6042 & 0.81 & 0.69 & 0.86 & 0.70 & 0.86\\
50 & 200 & -0.0131 & 0.1082 & 0.1083 & 0.4655 & 0.83 & 0.53 & 0.88 & 0.54 & 0.88\\
100 & 50 & -0.0032 & 0.3631 & 0.3631 & 0.7691 & 0.78 & 0.87 & 0.82 & 0.88 & 0.83\\
100 & 100 & -0.0046 & 0.1899 & 0.1900 & 0.5725 & 0.80 & 0.67 & 0.85 & 0.67 & 0.85\\
100 & 200 & -0.0065 & 0.1035 & 0.1035 & 0.4284 & 0.81 & 0.50 & 0.87 & 0.51 & 0.87\\
200 & 50 & -0.0024 & 0.3703 & 0.3703 & 0.7546 & 0.76 & 0.86 & 0.80 & 0.87 & 0.81\\
200 & 100 & 0.0018 & 0.1815 & 0.1815 & 0.5549 & 0.80 & 0.65 & 0.85 & 0.66 & 0.85\\
200 & 200 & -0.0036 & 0.0938 & 0.0938 & 0.4071 & 0.81 & 0.49 & 0.87 & 0.49 & 0.87\\
\hline
\end{tabular}
\begin{tablenotes}
 \item[\hskip -\fontdimen 2 \font]Note: ``Av. R. CI'', ``Av. R. CI QS ker.'' and ``Av. R. CI Parzen ker.'' (respectively ``Cov. CI'', ``Cov. CI QS ker.'' and ``Cov. CI Parzen ker.'') stand for the average radius (respectively, coverage) of the 95\% confidence intervals computed using, respectively, the standard estimator, the HAC estimator with quadratic spectral kernel and the HAC estimator with Parzen kernel of the covariance matrix.
\end{tablenotes}
\end{center}
\caption{Results for the estimator of the AME $\widehat{\Delta}$ with $J=2$} 
\label{tab.panelJ2}
\end{table}

  \end{landscape}
  \clearpage
}

\section{Empirical application} \label{sec.emp}
To illustrate our methodology, we revisit \cite{voigtlander2014skill}. This paper considers a panel data with $N=313$ sectors of the U.S. economy observed yearly from 1958 to 2005, that is $T=48$ dates. It investigates if increasing skill intensity of the economy is a cause of increasing wage inequality in the United States. To this end, \cite{voigtlander2014skill} builds an input skill intensity measure $\sigma_{i,t}$ and evaluate its effect on $\ln(w_{L,i,t}/w_{H,i,t})$, the logarithm of the ratio of the average wage of low-skilled workers in sector $i$ at time $t$, $w_{L,i,t}$, over the average wage of high-skilled workers in that sector at the same date. The goal is to assess if skill intensity in a given sector causes wage inequality in that sector. 

This data has recently been analyzed by \cite{yin2021focused} and \cite{juodis2022regularization} using common correlated effects approaches. We use the dataset of \cite{yin2021focused} available at \url{https://www.tandfonline.com/doi/abs/10.1080/07350015.2019.1623044?casa_token=Jmzhpt-330cAAAAA:MrhnmhaKnNDF1uWAl6jwO09Nz6S8Mx1oONJ7OHWmdLEagE0TeZ8pL60jSYL97XcXRjGwPaqYpkY}, it corresponds to a balanced version of the original data of \cite{voigtlander2014skill}, where some sectors with missing data have been deleted.

We let $y_{it}=\ln(w_{L,i,t}/w_{H,i,t})$ and $d_{it}=\ln(\sigma_{i,t})$. For the panel $\{x_{\ell t}\}_{\ell,t}$ we use 7 variables  for each sector at each date including the logarithm of the ratio of high skilled workers over low-skilled workers, capital equipment per worker. These corresponds to the control variables used in \cite{yin2021focused} and \cite{juodis2022regularization}.  This leads us to $L= 7\times 313=2191$ in the panel $\{x_{\ell,t}\}_{\ell,t}$. The control variables $c_{it}$ correspond to these $7$ variables along with a constant. 

Table \ref{tab.delta} reports the estimates and confidence intervals of $\widehat{\Delta}$ for $J=1,2,3$. The various confidence intervals are built as in the simulations and the growth ratio estimator finds one factor. The estimates and associated confidence intervals are not very sensitive to the choice of $J$. The results are significant and the coefficient is negative: increasing skill intensity increases wage inequality. Our estimates are much more negative than that of \cite{yin2021focused} (ranging from $-0.73$ to $-0.59$) and 
 \cite{juodis2022regularization} (ranging from $-0.99$ to $-0.59$).

\begin{table}
\centering
\begin{tabular}{|c|c|c|c|}
\hline
$J$& 1 & 2 &3\\
\hline
$\widehat{\Delta}$ & -1.71 & -1.78 & -1.75\\ 
CI & [-2.08, -1.34] & [-2.15,-1.41] & [-2.18,-1.31]\\
CI QS ker. & [-2.09, -1.33] & [-2.15,-1.41] & [-2.38,-1.11]\\
CI Parzen ker.& [-2.09, -1.33] & [-2.15,-1.41] & [-2.31,-1.18]\\
\hline
\end{tabular}
\begin{tablenotes}
 \item[\hskip -\fontdimen 2 \font]Note: ` ``CI'', ``CI QS ker.'' and ``CI Parzen ker.''' stand for the 95\% confidence intervals computed using, respectively, the standard estimator, the HAC estimator with quadratic spectral kernel and the HAC estimator with Parzen kernel of the covariance matrix.
\end{tablenotes}
\caption{Estimates and confidence intervals of $\widehat{\Delta}$ for $J=1,2,3$.}
\label{tab.delta}

\end{table}

\begin{figure}
\includegraphics[width=\textwidth]{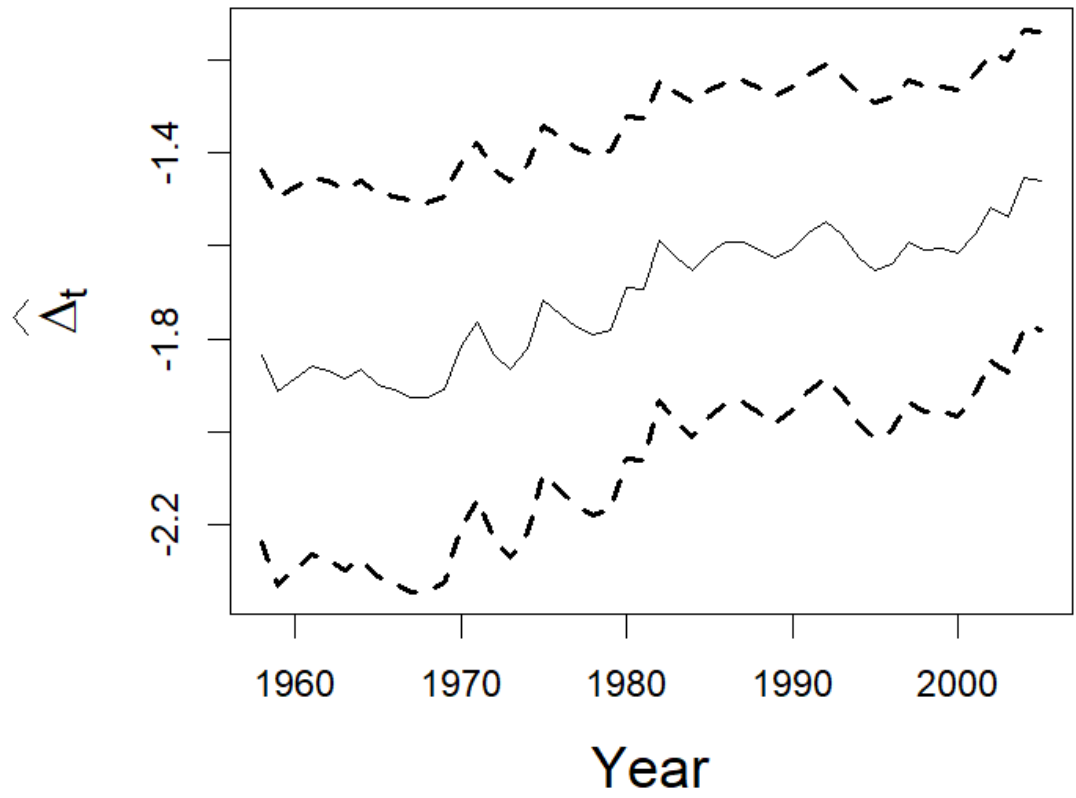}
\caption{Estimates (solid line) and pointwise confidence intervals (dashed lines) of the time-specific average marginal effects}
\label{fig.1}
\end{figure}

Next, we explore the time trend of the time-specific average marginal effect $\widehat{\Delta}_t$. In Figure \ref{fig.1}, we plot $\widehat{\Delta}_t$ along with its (pointwise) confidence intervals estimated with $J=1$. We see that the average marginal effect of skill intensity tends to increase over time, suggesting that the wage premium of skilled workers (for a fixed value of the skill intensity measure) is decreasing with time.

\section{Conclusion} \label{sec.ccl}In this paper, we present a novel factor model for potential outcomes, outlining the methodology for estimating various key parameters within the model. Notably, our approach accommodates both temporal and unit-specific variations in treatment effects. We further establish the asymptotic distribution for several pivotal estimators, paving the way for statistical inference. To enhance the simplicity of inference procedures, the development of a bootstrap technique for constructing confidence intervals across a broad spectrum of estimators in our model would be beneficial. This avenue remains open for exploration in future research endeavors. Another interesting extension is the case where the controls are high-dimensional and lasso is used for estimation. \cite{fan2022we} demonstrates the importance of such high-dimensional controls in counterfactual analysis with factor models and a binary treatment.

\if1\blind
{
\section*{Acknowledgments}
Jad Beyhum gratefully acknowledges financial support from the Research Fund KU Leuven through the grant STG/23/014. He thanks Andrii Babii, Geert Dhaene and Jonas Striaukas for useful comments.
} \fi

\bibliographystyle{agsm}
\bibliography{cfm}
\appendix 
\section*{Appendix}
\renewcommand\thesection{\Alph{section}}

\section{Preliminaries} We introduce further notation which will prove useful in the current section of the appendix. Let
\begin{align*}
W_i=(w_{i1}^\top,\dots,w_{iT}^\top)^\top,&\quad
\widehat{W}_i= (\widehat{w}_{i1}^\top,\dots,\widehat{w}_{iT}^\top)^\top
\end{align*}
be matrices of true and estimated regressors.  We also define $y_i=(y_{i1},\dots,y_{iT})^\top$, $c_i=(c_{i1},\dots,c_{iT})^\top$ and $u_i=(u_{i1},\dots,u_{iT})^\top$.
The following equality, corresponding to (13) in \cite{bai2023approximate},
\begin{equation}\label{decomp}
\widehat{F}-FH=\left(\frac{F\Lambda^\top E^\top \widehat{F} }{NT}+ \frac{E \Lambda F^\top \widehat F }{NT}+ \frac{ E E^\top \widehat{F}}{NT}\right)\left(\widehat{D}^2\right)^{-1},
\end{equation}
will be useful in the proofs.
Recall also that $\delta_{LT}=\sqrt{\min(L,T)}$.

\section{Useful lemmas on the factors} We prove some useful lemmas on the estimated factors. 
 \begin{Lemma}\label{lmm.bai}
 Let Assumptions \ref{A1}-\ref{A3} hold. We have 
 \begin{itemize}
 \item[(i)] $\frac1T \| \widehat{F}-F\widehat{H}\|_2^2 = O_P\left(\frac1L+\frac{1}{T^2}\right) $.
 \item[(ii)] $\widehat{D}^2\xrightarrow[T,L\to\infty]{\P}  D^2$.
 \item[(iii)] $\frac{\widehat{F}^\top F}{T}\xrightarrow{\P} Q$.
 \item[(iv)] $\widehat{H}\xrightarrow[T,L\to\infty]{\P} Q^{-1}.$
 \item[(v)] $\frac{\Lambda^\top E^\top \widehat{F}  }{LT}= O_P\left(\frac{1}{\sqrt{LT}}+ \frac1L\right)$.
 \item[(vi)] $\|E^\top \widehat{F}\|_2^2=O_P(LT)$.
 \item[(vii)] $\widehat{\Psi}\xrightarrow[T,L\to\infty]{\P}\Psi.$
\end{itemize}
\end{Lemma}
\begin{Proof} Statements (i), (ii), (iii), (iv) and (vi) correspond to the results of Proposition 1 and Lemmas 1, 2 and A.1 in \cite{bai2023approximate}. Statement (v) is shown in the proof of Lemma 3(i) in \cite{bai2023approximate}. Finally, (vii) is a direct consequence of (iv).
\end{Proof}

\begin{Lemma}\label{lm.big}
 Let Assumptions \ref{A1}-\ref{A4} hold. We have 
 \begin{itemize}
 \item[(i)] 
$\frac1T \| \widehat{W}_i-W_i\widehat{\Psi}\|_2^2 = O_P\left(\frac1L+\frac{1}{T^2}\right) $.
\item[(ii)] $\frac1T \| \widehat{Z}_i-Z_i\widehat{\Psi}\|_2^2 = O_P\left(\frac1L+\frac{1}{T^2}\right) $.
\item[(iii)] $\frac1T\sum_{t=1}^T  (\widehat{z}_{it}-\widehat{\Psi} z_{it})= O_P\left(\delta_{LT}^{-2}\right) $.
\item[(iv)] $\frac1TW_i^\top (\widehat{W}_i-W_i\widehat{\Psi}) = O_P\left(\delta_{LT}^{-2}\right) $.
\item[(v)] $\frac1T\widehat{W}_i^\top (\widehat{W}_i-W_i\widehat{\Psi}) = O_P\left(\delta_{LT}^{-2}\right) $.
\item[(vi)] $\frac1Tu_i^\top (\widehat{W}_i-W_i\widehat{\Psi}) = O_P\left(\delta_{LT}^{-2}\right) $.
\item[(vii)] $\frac1T\widehat{W}_i^\top\widehat{W}_i -\frac1T\widehat{\Psi}^\top W_i^\top W_i\widehat{\Psi} =o_P(1)$.
\item[(viii)] $ \frac{1}{\sqrt{N}} \sum_{i=1}^N (\widehat{\Psi}^{-1}\gamma_i)^\top (\widehat{z}_{it}-\widehat{\Psi}^\top z_{it})=  \frac{\sqrt{N}}{\sqrt{L}}\frac{1}{\sqrt{L}} \sum_{\ell=1}^L\left(\frac1N \sum_{i=1}^N \gamma_i^\top b_{i\ell t}\right)+O_P(\sqrt{N}/\delta_{LT}^2). $
\end{itemize}
\end{Lemma}
\begin{Proof}
Let us first show (i). The result follows from
\begin{align*}  
\left\|  \widehat{W}_i-W_i\widehat{\Psi} \right\|_2^2&=  \sum_{j=1}^J \sum_{t=1}^T \| \varphi_j(d_{it}) (f_t-\widehat{H}\widehat{f}_t)\|^2_2\\
&\le  \sum_{j=1}^J \sum_{t=1}^T |\varphi_j(d_{it})|^2\|  \widehat{f}_t-Hf_t\|^2_2\\
&\le C  \sum_{j=1}^J \sum_{t=1}^T\|  f_t-\widehat{H}\widehat{f}_t\|^2_2=  JC  \| \widehat{F}-F\widehat{H}\|_2^2=O_P\left(\frac{T}{L}+\frac{1}{T}\right),
\end{align*}
where the second inequality follows from Assumption \ref{A4} (ii) and we used Lemma \ref{lmm.bai} (i). 
The proof of (ii) is similar to that of (i), and is, therefore, omitted.

Then, we consider (iii). First, notice that 
 \begin{equation}\label{sumj}\left\|\frac1T\sum_{t=1}^T  (\widehat{z}_{it}-\widehat{\Psi}^\top z_{it})\right\|_2^2=\sum_{j=1}^J\|\ell_{j}^\top (\widehat{F}-F\widehat{H})\|_2^2,\end{equation}
where $\ell_{j} =(\varphi_j'(d_{i1}),\dots,\varphi_j'(d_{iT}))^\top$ for $j=1,\dots,J$. By \eqref{decomp}, we have 
\begin{align}\label{decompsum}
\ell_{j}^\top (\widehat{F}-FH)&=  \ell_{j}^\top\left(\frac{F\Lambda^\top E^\top\widehat{F} }{LT}+ \frac{E \Lambda F^\top \widehat F }{LT}+ \frac{ E E^\top \widehat{F}}{LT}\right)\left(\widehat{D}^2\right)^{-1}
\end{align}
Let us now bound each term.  We have first

\begin{equation}\label{3bounds}
\begin{aligned}
 \left\|\ell_{j}^\top\frac{F\Lambda^\top E^\top \widehat{F} }{LT}\right\|_2 &\le \|\ell_j\|_2\|F\|_2\frac{ \|\Lambda^\top E^\top \widehat{F} \|_2}{LT}=O_P\left(\sqrt{\frac{T}{L}}+ \frac{T}{L}\right).\\
\left\|\ell_{j}^\top\frac{E \Lambda F^\top \widehat F }{LT}\right\|_2 &\le \|\ell_j^\top E\Lambda \|_2 \|F\|_2\frac{\| \widehat{F} \|_2}{LT} =O_P\left(\sqrt{\frac{T}{L}}\right).\\
 \left\|\ell_{j}^\top\frac{ E E^\top \widehat{F}}{LT}\right\|_2 &\le \|\ell_{j}\|_2\|E\|_{op}  \frac{ \|E^\top \widehat{F}\|_2}{LT}=O_P\left(\sqrt{\frac{\max(L,T)}{L}}\right),
\end{aligned}
\end{equation}
where we used Lemma \ref{lmm.bai} (v) and (vi) and Assumptions \ref{A1} (ii) (d), \ref{A2} (i),  and \ref{A4} (iii) (a), (b) to obtain the bounds in \eqref{3bounds}. By \eqref{decompsum}, \eqref{3bounds} and Lemma \ref{lmm.bai} (ii), we get $T^{-1}\|\ell_{j}^\top (\widehat{F}-FH)\|_2^2=O_P(\delta_{LT}^{-2}).$ The result then follows from \eqref{sumj}.

Next, we show (iv). First notice that 
\begin{equation}\label{sumjj}
\begin{aligned}
&\frac1T \|W_i^\top (W_i\widehat{\Psi} -\widehat{W}_i)\|_2^2\\
&=   \left(\sum_{j,j'=1}^J \left\| \sum_{t=1}^T\varphi_{j'}(d_{it})f_t \varphi_j(d_{it}) (\widehat{f}_t-\widehat{H}f_t)^\top\right\|^2_2\right)+\left\|\sum_{t=1}^T c_{it}(\widehat{f}_t-\widehat{H}f_t)^\top\right\|_2^2\\
&=   \sum_{j,j'=1}^J \left\| \ell_{jj'}^\top(\widehat{F}-F\widehat{H})\right\|^2_2+\left\|c_i^\top (\widehat{F}-F\widehat{H})\right\|_2^2,
\end{aligned}
\end{equation}
where $\ell_{jj'}=(\varphi_j(d_{i1})\varphi_{j'}(d_{i1})f_1,\dots,\varphi_j(d_{iT})\varphi_{j'}(d_{iT})f_T)^\top$. 
By \eqref{decomp}, it holds that
\begin{equation}\label{decompsum2}
  \ell_{jj'}^\top(\widehat{F}-F\widehat{H})
 =  \ell_{jj'}^\top \left(\frac{F\Lambda^\top E^\top\widehat{F} }{LT}+ \frac{E \Lambda F^\top \widehat F }{LT}+ \frac{ E E^\top \widehat{F}}{LT}\right)\left(\widehat{D}^2\right)^{-1}.
\end{equation}
We bound each term. It holds that
\begin{equation}\label{3bounds2}
\begin{aligned}
 \left\|\ell_{jj'}^\top\frac{F\Lambda^\top E^\top \widehat{F} }{LT}\right\|_2 &\le \|\ell_{jj'}\|_2\|F\|_2\frac{ \|\Lambda^\top E^\top \widehat{F} \|_2}{LT}=O_P\left(\sqrt{\frac{T}{L}}+ \frac{T}{L}\right).\\
\left\|\ell_{jj'}^\top\frac{E \Lambda F^\top \widehat F }{LT}\right\|_2 &\le \|\ell_{jj'}^\top E\Lambda \|_2 \|F\|_2\frac{\| \widehat{F} \|_2}{LT} =O_P\left(\sqrt{\frac{T}{L}}\right).\\
 \left\|\ell_{jj'}^\top\frac{ E E^\top \widehat{F}}{LT}\right\|_2 &\le \|\ell_{jj'}\|_2\|E\|_{op}  \frac{ \|E^\top \widehat{F}\|_2}{LT}=O_P\left(\sqrt{\frac{\max(L,T)}{L}}\right),
\end{aligned}
\end{equation}
where we used Lemma \ref{lmm.bai} (v) and (vi) and Assumptions \ref{A1} (ii) (d), \ref{A2} (i),  and \ref{A4} (iii) (c), (d) to obtain the bounds in \eqref{3bounds2}. By \eqref{decompsum2}, \eqref{3bounds2} and Lemma \ref{lmm.bai} (ii), we get \begin{equation} \label{2912} \frac1T \|\ell_{jj'}^\top (\widehat{F}-FH)\|_2^2=O_P(\delta_{LT}^{-2}).\end{equation}
Let us now bound$\left\|c_i^\top (\widehat{F}-F\widehat{H})\right\|_2^2=O_P(\delta_{LT}^{-2})$. 
By \eqref{decomp}, we have
\begin{equation}\label{decompsum22}
  c_i^\top(\widehat{F}-F\widehat{H})
 =  c_i^\top \left(\frac{F\Lambda^\top E^\top\widehat{F} }{LT}+ \frac{E \Lambda F^\top \widehat F }{LT}+ \frac{ E E^\top \widehat{F}}{LT}\right)\left(\widehat{D}^2\right)^{-1}.
\end{equation}
We bound each term.
\begin{equation}\label{3bounds22}
\begin{aligned}
 \left\|c_i^\top\frac{F\Lambda^\top E^\top \widehat{F} }{LT}\right\|_2 &\le \|c_i\|_2\|F\|_2\frac{ \|\Lambda^\top E^\top \widehat{F} \|_2}{LT}=O_P\left(\sqrt{\frac{T}{L}}+ \frac{T}{L}\right).\\
\left\|c_i^\top\frac{E \Lambda F^\top \widehat F }{LT}\right\|_2 &\le \|c_i^\top E\Lambda \|_2 \|F\|_2\frac{\| \widehat{F} \|_2}{LT} =O_P\left(\sqrt{\frac{T}{L}}\right).\\
 \left\|c_i^\top\frac{ E E^\top \widehat{F}}{LT}\right\|_2 &\le \|c_i\|_2\|E\|_{op}  \frac{ \|E^\top \widehat{F}\|_2}{LT}=O_P\left(\sqrt{\frac{\max(L,T)}{L}}\right),
\end{aligned}
\end{equation}
where we used Lemma \ref{lmm.bai} (v) and (vi) and Assumptions \ref{A1} (ii) (d), \ref{A2} (i),  and \ref{A4} (iii) (b), (c) to obtain the bounds in \eqref{3bounds22}. By \eqref{decompsum22}, \eqref{3bounds22} and Lemma \ref{lmm.bai} (ii), we get \begin{equation} \label{29122} \frac1T \|c_i^\top (\widehat{F}-FH)\|_2^2=O_P(\delta_{LT}^{-2}).\end{equation}
The result then follows from \eqref{2912}, \eqref{29122} and \eqref{sumjj}.
The proof of (v) is a direct consequence of statements (i) and (iv) and it is therefore omitted.

Then, we prove (vi). First, notice that 
\begin{equation}\label{sumjjj}
\begin{aligned}
\frac1T \|u_{i}^\top (W_i\widehat{\Psi} -\widehat{W}_i)\|_2^2&=   \sum_{j}^J \left\| \sum_{t=1}^T\varphi_{j}(d_{it})u_{it} (\widehat{f}_t-Hf_t)^\top\right\|^2_2\\
&=  \sum_{j=1}^J \left\| g_j^\top(\widehat{F}-F\widehat{H})\right\|^2_2
\end{aligned}
\end{equation}
where $g_{j}=(\varphi_j(d_{i1})u_{i1},\dots, \varphi_{j}(d_{it})u_{iT})^\top$. 
By \eqref{decomp}, we have 
\begin{equation}\label{decompsum3}
g_j^\top(\widehat{F}-F\widehat{H})\\
 =g_j^\top \left(\frac{F\Lambda^\top E^\top\widehat{F} }{LT}+ \frac{E \Lambda F^\top \widehat F }{LT}+ \frac{ E E^\top \widehat{F}}{LT}\right)\left(\widehat{D}^2\right)^{-1}.
\end{equation}
Let us bound each term:
\begin{equation}\label{3bounds3}
\begin{aligned}
 \left\|g_j^\top\frac{F\Lambda^\top E^\top \widehat{F} }{LT}\right\|_2 &\le \|g_j\|_2\|F\|_2\frac{ \|\Lambda^\top E^\top \widehat{F} \|_2}{LT}=O_P\left(\sqrt{\frac{T}{L}}+ \frac{T}{L}\right).\\
\left\|g_j^\top\frac{E \Lambda F^\top \widehat F }{LT}\right\|_2 &\le \|g_j^\top E\Lambda \|_2 \|F\|_2\frac{\| \widehat{F} \|_2}{LT} =O_P\left(\sqrt{\frac{T}{L}}\right).\\
 \left\|g_j^\top\frac{ E E^\top \widehat{F}}{LT}\right\|_2 &\le \|g_j\|_2\|E\|_{op}  \frac{ \|E^\top \widehat{F}\|_2}{LT}=O_P\left(\sqrt{\frac{\max(L,T)}{L}}\right),
\end{aligned}
\end{equation}
where we used Lemma \ref{lmm.bai} (v) and (vi) and Assumptions \ref{A1} (ii) (d), \ref{A2} (i),  and \ref{A4} (iii) (e), (f) to obtain the bounds in \eqref{3bounds3}. By \eqref{decompsum3}, \eqref{3bounds3} and Lemma \ref{lmm.bai} (ii), we get $T^{-1}\|g_{j}^\top (\widehat{F}-FH)\|_2^2=O_P(\delta_{LT}^{-2}).$ The result then follows from \eqref{sumjjj}. Statement (vii) is a direct consequence of (i) and its proof is therefore omitted.

Finally, we prove (viii). Following the arguments of the proof of equation (20) in \cite{bai2023approximate}, we have  $$\sqrt{L} (\widehat{f}_t-\widehat{H}^\top f_t)=\frac{1}{\sqrt{L}} \widehat{H}^\top \Sigma_\Lambda^{-1}\sum_{\ell=1}^L \lambda_\ell e_{\ell t}+O_P(\sqrt{L}/\delta_{LT}^2).$$ Using the definitions of $\widehat{\Psi}$ and $b_{i\ell t}$ and 
Assumption \ref{A4} (ii)., we obtain 
$$ \sqrt{L} (\widehat{z}_{it}-\widehat{\Psi}^\top z_{it}) = \frac{1}{\sqrt{L}} \widehat{H}^\top \sum_{\ell=1}^L b_{i\ell t} + O_P(\sqrt{L}/\delta_{LT}^2),$$
which directly yields the result.
\end{Proof}

\section{Proof of Theorem \ref{AN}}
Note that $y_i= \widehat{W}_i \widehat{\Psi}^{-1} \gamma_i + (W_i\widehat{\Psi} -\widehat{W}_i)\widehat{\Psi}^{-1} \gamma_i+u_i.$
Plugging this in \eqref{defhatg}, we obtain
\begin{align*}
 \widehat{\gamma}_i&= (\widehat{W}_i^\top \widehat{W}_i)^{-1} \widehat{W}_i^\top\left(\widehat{W}_i \widehat{\Psi}^{-1} \gamma_i + (W_i\widehat{\Psi} -\widehat{W}_i)\widehat{\Psi}^{-1} \gamma_i+u_i\right).
\end{align*}
This leads to 
\begin{align*}  \sqrt{T}(\widehat{\gamma}_i- \widehat{\Psi}^{-1}  \gamma_i) &=  \sqrt{T}(\widehat{W}_i^\top \widehat{W}_i)^{-1} \widehat{W}_i^\top (W_i\widehat{\Psi} -\widehat{W}_i)\widehat{\Psi}^{-1} \gamma_i +  \sqrt{T}(\widehat{W}_i^\top \widehat{W}_i)^{-1} \widehat{W}_i^\top u_i \\
&= \sqrt{T}(\widehat{W}_i^\top \widehat{W}_i)^{-1} \widehat{W}_i^\top u_i+ O_P\left(\frac{\sqrt{T}}{\delta_{LT}^{2}}\right),\end{align*}
where we used that $\sqrt{T}(\widehat{W}_i^\top \widehat{W}_i)^{-1} \widehat{W}_i^\top (W_i\widehat{\Psi} -\widehat{W}_i)\widehat{\Psi}^{-1} \gamma_i=O_P\left(\sqrt{T}/\delta_{LT}^{2}\right)$ by Lemma \ref{lmm.bai} (vii) and \ref{lm.big} (v) and (vii) and Assumption \ref{A4} (v). Using Lemmas \ref{lmm.bai} (vii) and \ref{lm.big} (vi) and (vii), we also have $$\sqrt{T}(\widehat{W}_i^\top \widehat{W}_i)^{-1} \widehat{W}_i^\top u_i -\sqrt{T}(\Psi^\top W_i^\top W_i \Psi)^{-1} \Psi^\top W_i^\top u_i= O_P\left(\frac{\sqrt{T}}{\delta_{LT}^{2}}\right),$$
which yields \begin{equation}\label{expgamma}\sqrt{T}(\widehat{\gamma}_i- \widehat{\Psi}^{-1}  \gamma_i)=\sqrt{T}(\Psi^\top\Sigma_{w_iw_i}\Psi)^{-1}\Psi^\top W_i^\top u_i+ O_P\left(\frac{\sqrt{T}}{\delta_{LT}^{2}}\right).\end{equation}
Next, recall that
$\widehat{\Delta}_i= \frac{1}{T} \sum_{t=1}^T \widehat{\gamma}_i^\top \widehat{z}_{it}.$ This yields
\begin{equation}\label{expDelta}
\begin{aligned}
\widehat{\Delta}_i-\Delta_i &= \frac{1}{T} \sum_{t=1}^T (\widehat{\gamma}_i-\widehat{\Psi}^{-1} \gamma_i)^\top (\widehat{z}_{it}-\widehat{\Psi}^\top z_{it})+\left( \frac{1}{T} \sum_{t=1}^T z_{it} \right)^\top\widehat{\Psi}^\top (\widehat{\gamma}_i-\widehat{\Psi}^{-1} \gamma_i) \\
&\quad +(\widehat{\Psi}^{-1} \gamma_i)^\top \frac{1}{T} \sum_{t=1}^T (\widehat{z}_{it}-\widehat{\Psi}^\top z_{it})+  (\widehat{\Psi}^{-1} \gamma_i)^\top \widehat{\Psi}^\top\left( \frac{1}{T} \sum_{t=1}^T v_{it} \right).
\end{aligned}
\end{equation}
By Lemma \ref{lm.big} (iii) and the fact that $\widehat{\gamma}_i- \widehat{\Psi}^{-1}  \gamma_i=O_P(1/\sqrt{T})$ (from \eqref{expgamma} and Assumption \ref{A4} (iv) and (v) (a), (b)), it holds that 
\begin{equation}\label{negdelta}
\begin{aligned}
(\widehat{\gamma}_i-\widehat{\Psi}^{-1} \gamma_i)^\top  \frac{1}{\sqrt{T}} \sum_{t=1}^T (\widehat{z}_{it}-\widehat{\Psi}^\top z_{it})&=O_P(\delta_{LT}^{-2});\\
 (\widehat{\Psi}^{-1} \gamma_i)^\top \frac{1}{\sqrt{T }} \sum_{t=1}^T (\widehat{z}_{it}-\widehat{\Psi}^\top z_{it})&= O_P\left(\frac{\sqrt{T}}{\delta_{LT}^{2}}\right).
\end{aligned}
\end{equation}
By \eqref{expgamma}, \eqref{expDelta} and \eqref{negdelta}, we get 
\begin{equation}\label{expandelta}
\begin{aligned}
&\sqrt{T}(\widehat{\Delta}_i-\Delta_i )\\
 &=\sqrt{T} \left( \frac{1}{T} \sum_{t=1}^T  z_t \right)^\top\widehat{\Psi}^\top (\Psi^\top\Sigma_{w_iw_i}\Psi)^{-1}\Psi^\top W_i^\top u_i +  (\widehat{\Psi}^{-1} \gamma_i)^\top \widehat{\Psi}^\top\left( \frac{1}{T} \sum_{t=1}^T v_{it} \right)+ O_P\left(\frac{\sqrt{T}}{\delta_{LT}^{2}}\right) \\
& =\sqrt{T}\E_t[z_{it}]^\top\Psi^\top (\Psi^\top\Sigma_{w_iw_i}\Psi)^{-1}\Psi^\top \left(\frac{1}{\sqrt{T}}\sum_{t=1}^Tw_{it}u_{it}\right)+\gamma_i^\top\left( \frac{1}{\sqrt{T}} \sum_{t=1}^T v_{it} \right)+ O_P\left(\frac{\sqrt{T}}{\delta_{LT}^{2}}\right)\\
&= \omega_i^\top \left(\frac{1}{\sqrt{T}}\sum_{t=1}^Th_{it} \right)+ O_P\left(\frac{\sqrt{T}}{\delta_{LT}^{2}}\right),
\end{aligned}
\end{equation}
where in the second equality, we used Lemma \ref{lmm.bai} (vii) and Assumption \ref{A4} (v) (c). By Assumption \ref{A4} (d), we obtain $\sqrt{T}(\widehat{\Delta}_i-\Delta_i)=\omega_i^\top \left(\frac{1}{\sqrt{T}}\sum_{t=1}^Th_{it} \right)+o_P(1)$ and the asymptotic distribution follows from Assumption \ref{A4} (v) (b).

\section{Proof of Theorem \ref{ANMG}}
\subsection*{Proof of (i)} 
From $\widehat{\Delta}_t=\frac1N\sum_{i=1}^N\widehat{\gamma}_i^\top \widehat{z}_{it} $, we have
\begin{equation}\label{0201}
\begin{aligned}
\sqrt{N}(\widehat{\Delta}_t-\Delta_t) &= \frac{1}{\sqrt{N}} \sum_{i=1}^N (\widehat{\gamma}_i-\widehat{\Psi}^{-1}\gamma_i)^\top \widehat{\Psi}^{\top} z_{it}  +\frac{1}{\sqrt{N}} \sum_{i=1}^N (\widehat{\gamma}_i-\widehat{\Psi}^{-1}\gamma_i)^\top (\widehat{z}_{it}-\widehat{\Psi}^\top z_{it}) \\
&\quad + \frac{1}{\sqrt{N}} \sum_{i=1}^N (\widehat{\Psi}^{-1}\gamma_i)^\top (\widehat{z}_{it}-\widehat{\Psi}^\top z_{it})  + \frac{1}{\sqrt{N}} \sum_{i=1}^N\gamma_i^\top z_{it}- \Delta_t.
\end{aligned}
\end{equation}
Notice that, by \eqref{expgamma}, 
\begin{equation}\label{02012}
\begin{aligned}
  \frac{1}{\sqrt{N}} \sum_{i=1}^N (\widehat{\gamma}_i-\widehat{\Psi}^{-1}\gamma_i)^\top \widehat{\Psi}^{\top} z_{it} &= \frac{1}{\sqrt{N}} \sum_{i=1}^N  (\Psi^\top\Sigma_{w_iw_i}\Psi)^{-1}\Psi^\top \left(\frac1T \sum_{t=1}^T w_{it} u_{it} \right)^\top z_{it}+ O_P\left(\frac{\sqrt{N}}{\delta_{LT}^{2}}\right)\\
  &=o_P(1),
\end{aligned}
\end{equation}
where we used Assumption \ref{A5} (i) (a) and (ii).
Then, remark that 
\begin{equation}\label{02013} \frac{1}{\sqrt{N}} \sum_{i=1}^N (\widehat{\gamma}_i-\widehat{\Psi}^{-1}\gamma_i)^\top (\widehat{z}_{it}-\widehat{\Psi}^\top z_{it})=o_P(1),\end{equation}
by Lemma \ref{lm.big} (viii) and \eqref{expgamma}.
Next, by Lemma \ref{lm.big} (viii) and Assumption \ref{A5} (i) (b) and (ii), we have 
\begin{equation}\label{02014}
\begin{aligned} \frac{1}{\sqrt{N}} \sum_{i=1}^N (\widehat{\Psi}^{-1}\gamma_i)^\top (\widehat{z}_{it}-\widehat{\Psi}^\top z_{it}) &= \sqrt{\frac{L}{N}}\frac{1}{\sqrt{L}}\sum_{\ell=1}^L \left(\frac1N \sum_{i=1}^N \gamma_i^\top  b_{i\ell t} \right) +o_P(1) \\
&= \sqrt{\frac{N}{L}}\frac{1}{\sqrt{L}}\sum_{\ell=1}^L q_{\ell t}+o_P(1)   \end{aligned}
\end{equation}
Combining \eqref{0201}, \eqref{02012}, \eqref{02013}, \eqref{02014}, we obtain
$$\sqrt{N} (\widehat{\Delta}_t-\Delta_t) =\sqrt{\frac{N}{L}}\left(\frac{1}{\sqrt{L}} \sum_{\ell=1}^L q_{\ell t} \right)+\frac{1}{\sqrt{N}}\sum_{i=1}^N \gamma_i^\top z_{it} -\Delta_t+o_P(1) .$$
The asymptotic distribution is a consequence of Assumption \ref{A5} (iii). 

\subsection*{Proof of (ii)} 
From $\widehat{\Delta}= \frac1N \sum_{i=1}^N \widehat{\Delta}_i$, we have
\begin{equation}\label{3012}
\begin{aligned}
\sqrt{N}(\widehat{\Delta}_i-\Delta) &=  \frac{1}{\sqrt{N}} \sum_{i=1}^N \widehat{\Delta}_i-\Delta_i + \frac{1}{\sqrt{N}} \sum_{i=1}^N\Delta_i - \Delta\\
&= \frac{1}{\sqrt{N}}\sum_{i=1}^N \E_i[z_{it}]^\top\Psi^\top (\Psi^\top\Sigma_{w_iw_i}\Psi)^{-1}\Psi^\top \left(\frac{1}{T}\sum_{t=1}^Tw_{it}u_{it}\right)\\
&\quad 
+\frac{1}{\sqrt{N}}\sum_{i=1}^N \gamma_i^\top\left( \frac{1}{T} \sum_{t=1}^T v_{it} \right) \\
&\quad +\frac{1}{\sqrt{N}} \sum_{i=1}^N\Delta_i - \Delta+O_P\left(\frac{\sqrt{N}}{\delta_{LT}^{2}}\right),
\end{aligned}
\end{equation}
where we used \eqref{expandelta} (which holds uniformly in i since Assumption \ref{A4} is imposed uniformly).
By Assumption \ref{A6} (i) (a), it holds that
\begin{equation}\label{30121}
 \frac{1}{\sqrt{N}}\sum_{i=1}^N \E_i[z_{it}]^\top\Psi^\top (\Psi^\top\Sigma_{w_iw_i}\Psi)^{-1}\Psi^\top \left(\frac{1}{T}\sum_{t=1}^Tw_{it}u_{it}\right)=O_P\left(\frac{1}{\sqrt{T}}\right).
\end{equation}
Moreover, note that
\begin{equation}\label{30122}
\begin{aligned}
&\frac{1}{\sqrt{N}} \sum_{i=1}^N \gamma_i^\top \left( \frac{1}{T} \sum_{t=1}^T v_{it} \right) \\
&=\frac{1}{\sqrt{N}}  \sum_{i=1}^N (\gamma_i-\gamma)^\top \left( \frac{1}{T} \sum_{t=1}^T v_{it} \right) + \frac{1}{\sqrt{N}} \sum_{i=1}^N \gamma^\top \left( \frac{1}{T} \sum_{t=1}^T v_{it} \right) \\
&=\frac{1}{\sqrt{N}} \sum_{i=1}^N  \gamma^\top \left( \frac{1}{T} \sum_{t=1}^T v_{it} \right)+O_P\left(\frac{1}{\sqrt{T}}\right),
\end{aligned}
\end{equation}
where we used Assumption\ref{A4} (i) (b). Additionally, remark that, for all $j\in[J]$, we have
\begin{align*}
&\frac{1}{NT}\sum_{i=1}^N  \sum_{t=1}^T\varphi_{j}'(d_{it})f_t-\E_t\left[\varphi_j'(d_{it})f_t\right]\\
&= \frac{1}{NT}\sum_{i=1}^N  \sum_{t=1}^T(\varphi_{j}'(d_{it})-\E_i[\varphi_j'(d_{it})])f_t+ \frac{1}{NT}\sum_{i=1}^N  \sum_{t=1}^T\E_i
[\varphi_j'(d_{it})]f_t-\E\left[\varphi_j'(d_{it})f_t\right]\\
&=\frac{1}{T}\sum_{t=1}^T\E_i
[\varphi_j'(d_{it})]f_t-\E\left[\varphi_j'(d_{it})f_t\right]+O_P\left(\frac{1}{\sqrt{NT}}\right),
\end{align*}
where we used Assumption \ref{A6} (i) (c). By definition of $v_{it}$ and $m_t$, this yields
\begin{equation}\label{30123}\sqrt{N}  \gamma^\top \left( \frac{1}{NT} \sum_{i=1}^N  \sum_{t=1}^T v_{it} \right)= \sqrt{N} \gamma^\top \left( \frac{1}{T}  \sum_{t=1}^T m_t \right)+O_P\left(\frac{1}{\sqrt{T}}\right).\end{equation}
Combining \eqref{3012}, \eqref{30121}, \eqref{30122} and \eqref{30123}, and using Assumption \ref{A4} (ii), we obtain 
 $$\sqrt{N}(\widehat{\Delta}-\Delta)=\sqrt{\frac{N}{T}} \gamma^\top\left(\frac{1}{\sqrt{T}}\sum_{t=1}^T m_{t} \right)+\frac{1}{\sqrt{N}}\sum_{i=1}^N \Delta_i-\Delta+o_P(1). $$
The asymptotic distribution then follows from Assumption \ref{A6} (iii).

\end{document}